\def\changesHilighted{false} %コメントオフ：投稿バージョン
\documentclass[dvipdfmx]{interact}

\usepackage{epstopdf}% To incorporate .eps illustrations using PDFLaTeX, etc.
\usepackage[caption=false]{subfig}% Support for small, `sub' figures and tables

\usepackage[numbers,sort&compress]{natbib}% Citation support using natbib.sty
\bibpunct[, ]{[}{]}{,}{n}{,}{,}% Citation support using natbib.sty
\makeatletter% @ becomes a letter
\def\NAT@def@citea{\def\@citea{\NAT@separator}}% Suppress spaces between citations using natbib.sty
\makeatother% @ becomes a symbol again

\theoremstyle{plain}% Theorem-like structures provided by amsthm.sty
\newtheorem{theorem}{Theorem}[section]

\theoremstyle{definition}

\newtheorem{problem}[theorem]{Problem}

\theoremstyle{remark}

\usepackage[dvipsnames]{xcolor}
\usepackage{mathtools}
\usepackage{svg}
\usepackage{amsmath}
\usepackage[normalem]{ulem}
\DeclarePairedDelimiter{\abs}{\lvert}{\rvert}
\newcommand{\norm}[1]{\lVert #1 \rVert}
\DeclareMathOperator*{\argmin}{argmin}
\DeclareMathOperator*{\argmax}{argmax}

\definecolor{dogwoodrose}{rgb}{0.84, 0.09, 0.41}
\definecolor{darkpastelgreen}{rgb}{0.01, 0.75, 0.24}

\newcommand{\add}[1]{\textcolor{\addcolor}{#1}}
\renewcommand{\add}[1]{#1}

\newcommand{\anna}[1]{\textcolor{MidnightBlue}{\ [藤岡：#1]}}
\newcommand{\masaki}[1]{\textcolor{dogwoodrose}{{\ [小蔵：#1]}}}

\newcommand{\waka}[1]{\textcolor{Gray}{\textbf{\ [若宮：#1]}}}

\newcommand{\ask}[1]{\textcolor{darkpastelgreen}{\textbf{\ [お願い：#1]}}}
\newcommand{\resolved}[1]{\textcolor{SeaGreen}{{\ [解決済：#1]}}}

\newcommand{\addd}[1]{\textcolor{\addcolor}{#1}}
\renewcommand{\addd}[1]{#1}

\newcommand{\del}[1]{\textcolor{red}{\sout{#1}}}
\renewcommand{\del}[1]{}

\newcommand{\adddd}[1]{\textcolor{\addcolor}{#1}}
\renewcommand{\adddd}[1]{#1}

\newcommand{\dell}[1]{\textcolor{red}{\sout{#1}}}
\renewcommand{\dell}[1]{}

\newcommand{\delwhole}[1]{{#1}}

\newcommand{\addddd}[1]{\textbf{\textcolor{\addcolor}{#1}}}

\newcommand{\delll}[1]{\textcolor{red}{\sout{#1}}}

\newcommand{\textbftemp}[1]{\textbf{#1}}

\ifnum\pdfstrcmp{\changesHilighted}{false}=0
    \renewcommand{\textbf}[1]{#1}
    \renewcommand{\anna}[1]{}
    \renewcommand{\masaki}[1]{}
    \renewcommand{\waka}[1]{}
    \renewcommand{\ask}[1]{}
    \renewcommand{\resolved}[1]{}
    \renewcommand{\delwhole}[1]{}
    \renewcommand{\delll}[1]{}
    \renewcommand{\addddd}[1]{#1}
\fi

\usepackage{url}

\begin{document}

% \articletype{ARTICLE TEMPLATE}% Specify the article type or omit as appropriate

\title{Shepherding Heterogeneous Flock with Model-Based Discrimination}

% モデルベースの話を入れる？
% Shepherding Heterogeneous Flock with Model Predictive ...
% MSCS ？
% Model-Based Shepherding Algorithm for Flocks with Unexpected Agents

\author{
% \name{A.~N. Author\textsuperscript{a}\thanks{CONTACT A.~N. Author. Email: latex.helpdesk@tandf.co.uk} and John Smith\textsuperscript{b}}
\name{Anna Fujioka\textsuperscript{a}, Masaki Ogura\textsuperscript{b} and Naoki Wakamiya\textsuperscript{b}}
\affil{\textsuperscript{a}School of Engineering and Science, Osaka University \textsuperscript{b}Graduate School of Information Science and Technology, Osaka University}
}

\maketitle

\begin{abstract}
% 群れをなす多数のエージェントを少数の外部エージェントがもたらす斥力を利用して目的地へ誘導する問題は，牧羊犬が羊群を誘導する行動からshepherding問題と呼ばれる．この問題は，飛行機事故を防ぐために滑走路から鳥を遠ざけるための誘導，海に流出した石油の回収，地図の作成のためのロボット群の開発など，様々な応用の可能性から注目を集めている．このshepherding問題について多くの研究が行われてきたが，それらの研究のほとんどにおいて，被誘導エージェントのダイナミクスは一様であると仮定されていた．しかしながら，現実の場面ではダイナミクスが一様であるとは限らない．そこで本報告では，誘導対象である通常のエージェントとは異なるダイナミクスを持つエージェントが混入した群れに対する誘導手法を提案する．この誘導においては，誘導対象エージェントの想定ダイナミクスからの乖離度に基づいた混入エージェント検出が重要な役割を果たす．この乖離度の計算には，想定する軌道と実際の軌道との差を用いる．本報告では予測軌道の定め方と混入エージェントの検出方法に関する四つの手法を提案する．提案手法の有効性と，被誘導エージェントのモデル化誤差に対するロバスト性の有無をシミュレーションにより示す．シミュレーションの結果，混入エージェントのダイナミクスに関わらず，従来手法よりも提案手法の方が多くの誘導対象エージェントを誘導したことが示された．
The problem of guiding a flock of agents to a destination by the repulsion forces exerted by a smaller number of external agents is called the shepherding problem. This problem has attracted attention due to its potential applications, including diverting birds away for preventing airplane accidents, recovering spilled oil in the ocean, and guiding a swarm of robots for mapping. Although there have been various studies on the shepherding problem, most of them place the uniformity assumption on the dynamics of agents to be guided. However, we can find various practical situations where this assumption does not necessarily hold. In this paper, we propose a shepherding method for a flock of agents consisting of normal agents to be guided and other variant agents. In this method, the shepherd discriminates normal and variant agents based on their behaviors' deviation from the one predicted by the potentially inaccurate model of the normal agents. As for the discrimination process, we propose two methods using static and dynamic thresholds.
\dell{The effectiveness of the proposed methods and their robustness to modeling errors of the normal agents are demonstrated through simulations. }
% \anna{削除ありがとうございます}
Our simulation results show that the proposed methods outperform a conventional method for various types of variant agents.
\end{abstract}


\delwhole{\color{dogwoodrose} 全般的なメモ
\begin{itemize}
    % \item 2022-03-22の気持ち：若宮先生へは金曜日か月曜日で良いであろう
    \item Deepl結果に対する差分は logAR2022\_2022-03-22\_10時03分.pdf
    % \item 藤岡さん色と区別するために差分の色をSepiaぽいのに戻しました．僕のスクリーン上で見やすいように少し薄くしました．
    % \item 作業している感を出すために若宮先生へはコメントをある程度残したバージョンを見せたいと思っています．このコメント自体はもちろん削除します．
    \item 投稿バージョンにするためにはmain.texの1, 2行目をいじります．精読のときは投稿バージョンにすると便利です．
    \item 編集の妨げになってきたので古いコメントを消しました（2022/03/30 14:30）
\end{itemize}}

% \delwhole{\color{dogwoodrose} Todoリスト
% \begin{itemize}
%     \item \done{日本人の論文を引用，櫻間先生？\url{https://ieeexplore.ieee.org/abstract/document/9099405}, \url{https://www.tandfonline.com/doi/abs/10.1080/00207721.2013.798442} \url{https://link.springer.com/article/10.1007/s10489-020-02034-2}
%     \url{https://www.jstage.jst.go.jp/article/iscie/31/1/31_21/_article/-char/ja/}}
%     \item \done{自動車？}
%     \item \done{他の shepherding research 論文引用？}
%     \item \done{問題設定の妥当性？\url{https://jglobal.jst.go.jp/detail?JGLOBAL_ID=202002289719782212}}
% \end{itemize}}

% \resolved{section titleがtitle caseかsentence caseか教えてください．またそれを原稿に反映してください．}\anna{どこでその情報を得られるかわかりませんでした．すみません．styファイルダウンロードの際の注意事項記載のinteractnlmsample.pdfとslackで送っていただいたURL確認しましたが，自分の力では見つけ出せませんでした．pdfではsentence case でしたので，一旦sentense case にしています．どこから情報を得られるか教えていただきますと幸いです．．}\masaki{interactnlmsample.pdfにおいてsentence caseなので，僕らもsentence caseにします（なので現状そのままでOK．}

\section{Introduction}
% 自律的に移動する複数のエージェントの群れを誘導することによって何らかのタスクを達成するシステムは，その応用の可能性から注目を集めている．応用の例として，バードストライクを防ぐために滑走路から鳥を遠ざけるための誘導\cite{gade2015herding}，海や川などに流出した石油の収集\cite{zahugi2013oil}，地図の作成のためのロボット群の制御\cite{kegeleirs2021swarm}などが挙げられる．このようなシステムについて，様々な群れに対する誘導手法が提案されてきた．群れの誘導は，誘引力に基づく誘導と斥力に基づく誘導の二つに分類することができる．しかしながら，この二つの誘導を比較すると，斥力に基づく誘導の方が誘引力に基づく誘導よりも群れ全体を誘導しやすいことが示されている\cite{goel2019leader}ため，本報告では斥力に基づく誘導に注目する．

\addd{The guidance and navigation of flocks of agents have several applications including} guiding birds away from runways \addd{for preventing} bird strikes~\cite{gade2015herding}, collecting oil spills in oceans and rivers~\cite{zahugi2013oil,Pashna2020}, and \addd{navigating} a swarm of robots for map creation~\cite{kegeleirs2021swarm} \adddd{and coverage~\cite{Gusrialdi2008}}. For such systems, a variety of guidance methods for flocks have been proposed in the literature. \addd{The flock} guidance methods \addd{available in the literature} can be mainly classified into the following two categories: attraction-based guidance and repulsion-based guidance. Motivated by a recent comparison~\cite{goel2019leader} of these two types of guidance methods \addd{suggesting} the potential superiority of the repulsion-based method over the attraction-based method, this \addd{paper} focuses on the repulsion-based guidance of flocks of agents.

% 斥力に基づく誘導手法の一つとして，牧羊犬が羊群を誘導する行動から着想を得たshepherdingと呼ばれる枠組みの誘導への期待があり，この誘導の達成を目的とするshepherding問題の研究が行われている\cite{long2020comprehensive}．この問題における目的は，群れをなす多数のエージェントを少数のエージェントが自身の斥力を用いて与えられた目的地まで誘導するような少数エージェントの機動法則の設計である．被誘導エージェントは，群れ内部の相互作用と誘導側エージェントから受ける斥力により移動する．群れ内部の相互作用として，多くの場合Reynolds~\cite{reynolds1987flocks}によって提案されたBoidモデルの分離，整列，結合の三種類が仮定される．

\addd{The} repulsion-based guidance framework for flocks called shepherding\addd{ \cite{long2020comprehensive}} is \addd{an emergent framework} inspired by the behavior of sheepdogs guiding a flock of sheep. \addd{Specifically, the shepherding problem refers to the} problem \addd{of designing} \addd{the movement law} of a small number of \addd{external steering} agents \addd{(called shepherds)} so that they can guide\addd{, with their repulsion force,} a larger number of agents \addd{(called sheep)} to a given destination. \addd{Consequently, in the course of the navigation by the shepherd agents, the} sheep agents move according to their \addd{inter-flock} interactions and the repulsive forces from the shepherds. As for the \addd{inter-flock} interactions, \addd{the following} three types of interactions \addd{in the Boid model~\cite{reynolds1987flocks}} are often assumed: separation, alignment, and attraction\resolved{ここはattractionが正しいですか？もしそうであれば更新おねがいします}.\anna{attractionです．直しました．}\masaki{thanks}

% このようなshepherding問題に対して，誘導側エージェントの様々な移動法則が提案されてきた．例えば，Vaghanら\cite{vaughan1998robot}は，群れの中心に近づくことで群れ全体を誘導する移動法則を提案し，ロボットがアヒルの群れを誘導する実験を通じて移動法則の有効性を示した．
% また，Str\"{o}mbomら\cite{strombom2014solving}は，実際の牧羊犬の振る舞いに基づいて，群れから離れた個体を群れに近づけるCollectingと，群れの中心に近づくDrivingの二つの手法を使い分けながら群れを誘導する移動法則を提案した．Tsunodaら\cite{tsunoda2018analysis}は，群れの中で最も目的地から離れた個体を誘導する移動法則を提案した．その移動法則がVaghanやStr\"{o}mbomの移動法則よりも高い誘導性能を有することを示した．その後Tsunodaら\cite{tsunoda2019analysis}は，文献\cite{tsunoda2018analysis}の移動法則における課題であった，群れの分断を防止するように改良した移動法則を提案した．
% Huら\cite{hu2020occlusion}は，群れの背後に回り込むことで群れを誘導する移動法則を提案し，移動法則の有効性をシミュレーションとロボットによる実験により示した．
% KoとZuazua~\cite{ko2020asymptotic}は，ゴール領域から脱出しようとする回避者の群れに対するフィードバック戦略を提案した．

\del{For this kind of shepherding problems, researchers have proposed various movement rules of the guiding agents.}\addd{In the literature of the shepherding problem, we can find several movement laws of shepherds for guiding the sheep agents under various problem settings.} For example, Vaghan et al.~\cite{vaughan1998robot} proposed a \adddd{shepherd's} movement law \addd{in which the shepherd agent accomplishes guidance by moving toward the center of the sheep flock}\del{that guides the entire flock by having the shepherd agent approach the center of the flock}, and demonstrated the law's effectiveness through \addd{robotic} experiments \addd{for guiding}\del{in which a robot guided} a flock of ducks. Str\"{o}mbom et al.~\cite{strombom2014solving} proposed a \addd{shepherd's} movement \del{rule that guides a flock}\addd{law in which the shepherd alternatively uses} \del{by using}\addd{the following} two different methods \del{based on}\addd{inspired by} the behavior of actual sheepdogs: collecting, which brings \addd{closer to the flock the} individuals away from the flock\del{ closer to the flock}, and driving, which brings \del{them closer}\addd{the whole flock} to \del{the center of the flock}\addd{the goal}. Tsunoda et al.~\cite{tsunoda2018analysis} proposed a \del{migration}\addd{shepherd's movement} law\add{, called the Farthest-Agent Targeting (FAT) method, in which the shepherd} \del{that }guides the\del{ individual in the flock that is} \add{sheep} farthest from the destination, and have shown that the proposed movement law can outperofrm the movement laws of Vaghan et al. and Str\"{o}mbom et al. The \add{same} authors have further shown \add{in their another work~\cite{tsunoda2019analysis}} an improved version of the FAT method by introducing a modification for preventing the scatterment of flocks. Hu et al.~\cite{hu2020occlusion} proposed a shepherding method in which the shepherd guides the flock by going behind the herd, and showed the effectiveness of the method by both simulations and robotic experiments. Ko and Zuazua~\cite{ko2020asymptotic} proposed a feedback-based shepherding method for a flock of agents trying to escape from a goal area.

% 多くの従来手法では，被誘導エージェントのダイナミクスは一様であると仮定され ている.しかしながら，現実の場面ではダイナミクスが一様であるとは限らない.例え ば，魚は捕食者から身を守るために群れを形成するが，個体の動きは一様でなく，周り の魚の密度や速度から影響を受ける \cite{hemelrijk2008self}.このような状況に対する手法の一つとして， Himo ら\cite{himo2022iterative} は誘導側エージェントに反応しない被誘導エージェントが存在する群れに 対する誘導手法を提案した.この誘導手法においては，誘導側エージェントは被誘導 エージェントのダイナミクスを識別できると仮定しているが，その仮定は現実の場面で は必ずしも成り立たない.例えば，緊急時の群衆の誘導の場面においては，人々は押し 合いや踏みつけなど予期しない行動を起こすことがあるが，これらの行動は個人の本能 や経験，周囲の行動などに基づくため，誰が予期しない行動をとるかを事前に知ること は難しい\cite{pan2007multi}.

\addd{A common practice in the literature of the shepherding problem is placing the uniformity assumption on the}\del{Most conventional methods assume a uniform} dynamics within the flock of \addd{sheep} agents \del{being guided}\addd{to be guided}. However, \add{this assumption does not necessarily hold true} in \add{several} practical \addd{scenarios}\del{situations, there is no guarantee that the dynamics are uniform}. For example, while fish form schools to protect themselves from predators, the \del{movement}{dynamics} of \del{these}\add{each} individual\del{s} is not \del{always}\add{necessarily} uniform~\cite{hemelrijk2008self}\del{affected by the density and speed of the fish around them}\resolved{この部分は相談しておくべきでした，すみません．以下について意見をお願いします．原文は「魚は捕食者から身を守るために群れを形成するが，個体の動きは一様でなく，周り の魚の密度や速度から影響を受ける」がダイナミクスの非一様性を主張する文として使われていましたが，これが必ずしも適切ではないと感じていました．「周りの魚の密度や速度から影響を受ける」はダイナミクスが一様であっても非一様であっても起こりうる話ですので，あまり関係無いような気がします．僕の魚に対する理解は，仮にダイナミクス（動き方のルール）が一様な魚群であっても，個々の魚は周りの魚の密度や速度から影響を受けるので，結果として動きは違って見える，というものです．たとえば群れの真ん中にいる魚と端っこにいる魚は，ダイナミクスは一緒でも，動き方は違う．一方で，たぶんダイナミクスが異なるというのは真なのだと思います．全ての魚が同一のダイナミクスを持っているというのは想像しづらいです．これらを踏まえてこの部分の表現は，日本語レベルでは「捕食者から身を守るなどの目的で群れされる魚群においてそれぞれの個体のダイナミクスは異なることが知られている[14]」みたいな感じにするのはどうでしょうか．}\anna{よく考えたらBoidモデルもダイナミクスは一様だけど周囲の状況によって動き異なりますね．文章賛成です．ありがとうございます．}\masaki{thanksアップデートしました}. \adddd{On the other hand, in the context of the swarm robotics, heterogeneity within the swarm can be found due to fluctuations in production processes~\cite{Sakurama2020,Sakurama2015} or by the intent of the operator of the swarm~\cite{Aotani2019}.} \add{In order to address the aforementioned gap between the literature and the practice,} Himo et al.~\cite{himo2022iterative} recently proposed a guidance method for a flock \del{in which there is a guided agent}\addd{containing agents} \del{that does not respond}\addd{not responding} to the \del{guiding}\addd{shepherd} agent. \addd{Although this work sheds light on the shepherding-type guidance of a heterogeneous flock, the}\del{Although this} guidance method \addd{still} assumes \del{that }the \del{guiding agent can identify the}\addd{shepherd's knowledge of the type of each sheep agent}\del{ agent}, \addd{thereby having a limited applicability}\del{this assumption does not necessarily hold true} in \add{some} practical situations. For example, in emergency crowd control situations, people may engage in unexpected behaviors such as pushing and trampling. However, it is difficult to know in advance who will behave unexpectedly because these behaviors are based on individual instincts, experiences, and the actions of those around them (see, e.g., \cite{pan2007multi,Fujita2019}).

% そこで本報告では，以下のようなshepherding問題を考える．群れが誘導対象のエージェントと異種のエージェントから構成される状況を考える．異種エージェントは誘導対象エージェントとは異なるダイナミクスを有するものとする．また，誘導側エージェントは誘導対象エージェントのダイナミクスに関する情報は与えられているが，異種エージェントに関する情報は与えられていないものとする．この問題における誘導側エージェントの目的は，自身の移動と斥力による誘導対象エージェント群の目的地までの誘導である．
Therefore, in this paper, we consider a problem of shepherding a \emph{heterogeneous} flock with a shepherd agent having \emph{no prior information on the type of each sheep}. We specifically consider a situation in which a flock consists of \addddd{the following two types of sheep agents:} normal sheep agents \delll{to be guided }and \waka{唐突．
ここまでの流れは，「Shepherdingの研究はホモ対象だった，けど現実の状況はヘテロ，ヘテロに対しては日茂が取り組んだけど事前知識を必要としていた」なので直接的には「本論文では事前知識を必要としないやり方を提案する」であって，これまで言及されてこなかった「異種は誘導しなくてもいい」がでてくるのは唐突に感じる．従来研究の前提（ホモ）に対してクレームしているのでここで新たにおいた前提（異種は誘導しなくてもいい）についてはそれなりに妥当な説明が求められる．（人の前提を否定しておきながら自分は自分に都合の良い前提をおいているように見える）}variant sheep agents\delll{ not necessarily to be guided}. \ask{書き直してみたので以下確認おねがいします}\anna{ありがとうございます．確認しました．}\addddd{As for the normal sheep agents, the shepherd is assumed to be given information about their dynamics. On the other hand, as for the variant sheep agents, we assume that their dynamics are different from those of the normal sheep and, furthermore, are unknown to the shepherd.}\delll{The variant sheep are assumed to have dynamics different from those of the normal sheep. It is further supposed that the shepherd is given information about the dynamics of the normal agents to be guided, while no information about the variant sheep is given to the shepherd.} \addddd{A major difference of our problem formulation from the one in~\cite{himo2022iterative} is that we allow the dynamics of the variant sheep to lack any of the alignment, separation, attraction, and repulsion from the shepherd. Therefore, the situation we consider in this paper includes the case in which the navigation of the variant sheep is essentially a difficult task. For this reason, in our problem formulation, we consider the navigation of only the normal sheep.} \delll{The}{\addddd{Hence, the} goal of the guidance by the shepherd is set to be guiding the group of \addddd{only the} normal agents to the destination area.}

% この問題に対して本報告では，誘導対象エージェントの数理モデルを利用した誘導側エージェントの移動法則を提案する．誘導対象エージェントの数理モデルに基づいて被誘導エージェントの軌道を予測する．予測軌道からの乖離度に基づき異種エージェントを検出する．検出された異種エージェントを除外する．Farthest-Agent Targeting 法\cite{tsunoda2019analysis}を適用することで誘導対象エージェントを目的地まで誘導する．

Because we assume that the shepherd has no prior knowledge on the type of respective sheep, the methodology presented in~\cite{himo2022iterative} is not directly applicable to the current problem setting. Therefore, in this paper, we propose a discrimination method in which the nominal model of the normal sheep agents is utilized. Specifically, the shepherd internally predicts the trajectory of sheep agents under the assumption that all the sheep are normal. The shepherd then computes the deviation of the trajectory of each sheep from its prediction for discriminating agents. Finally, for agents that are \adddd{not} discriminated to be \dell{normal}\adddd{variant}%
% \resolved{これで正しいですか？確認お願いします}\anna{ありがとうございます．正しいです．}
, the shepherd agent applies the FAT method to guide them to the destination area.
\adddd{We remark that, due to this model-based characteristics of the discrimination process, the proposed shepherding method can be considered to be an application of the framework called Model Predictive Control (MPC) in the systems and control theory~\cite{Mayne2014}. Although there exist several works on the Model Predictive Control of heterogeneous multi-agent systems (see, e.g., \cite{Okajima2014,Kawamura2019}), their direct application to the current problem is not necessarily realistic due to the high nonlinearity in the Boid model. For this reason, in this paper we develop a novel shepherding algorithm and, furthermore, aim to establish its effectiveness via extensive numerial simulations.}
% 本報告の構成は以下の通りである．第2章では本報告で取り扱うshepherding問題を定式化する．第3章では本報告で提案する予測軌道に基づいた誘導手法について述べる．第4章ではシミュレーションにより既存手法の一つであるFarthest-Agent Targeting 法\cite{tsunoda2019analysis}と比較することにより，提案手法の有効性を評価する．異種エージェントの種類やその個体数を変化させながら誘導成功率を評価する．また，被誘導エージェントのモデル化誤差に対するアルゴリズムのロバスト性を評価する．第5章では，本報告のまとめと今後の課題を述べる．

This paper is organized as follows. In Section~\ref{sc:setting}, we formulate the shepherding problem studied in this paper. In Section~\ref{sc:virtual_method}, we describe the proposed guidance method based on model-based discrimination. In Section~\ref{sc:sml_virtual}, we evaluate the effectiveness of the proposed method through comparison with the FAT method by numerical simulations. We specifically evaluate the dependence of the guidance success rate on the type and the number of variant agents. %
% We also evaluate the robustness of the proposed method against modeling errors of the agents to be guided.
% \masaki{藤岡さん：これは本論文ではやっていないですよね？}\anna{ロバスト性やってないですね．．コメントアウトしました}\masaki{thanks}
In Section~\ref{sc:conclusion}, we conclude the paper and discuss future research directions.

\section{Problem statement}\label{sc:setting}
% 本報告で取り扱うshepherding問題を定式化する．二次元平面$\mathbb R^2$におけるマルチエージェントシステムを考える．システムは被誘導側のエージェント$N$体と誘導を行うエージェント一体から構成されるとする．本報告ではshepherding問題における慣例\cite{long2020comprehensive}に従い, 誘導されるエージェントを羊, 誘導するエージェントを牧羊犬と呼ぶ．羊は\ref{subsc:sheep}節で述べるモデルに従って群れ行動と牧羊犬からの回避行動をとる．牧羊犬の目的は, 自身の移動を通じて羊を所定の目的地$G\subset \mathbb R^2$へ誘導することである．目的地は中心$x_g\in \mathbb R^2$, 半径$R_g > 0$の開円板の領域とする．

In this section, we formulate the shepherding problem studied in this paper. \waka{N agentsが"to be guided"と書いてあるため異種（not necessarily to be guided）は含まないような気がしてしまうので，（先を読めば分かるけど）「含む」と明記した方がいい．}Let us consider a multi-agent system on the two-dimensional plane~$\mathbb R^2$. The multi-agent system consists of $N$ agents\addddd{, each of which is either} to be guided \addddd{or not to be guided}, and one steering agent performing navigation. Following the convention in the literature~\cite{long2020comprehensive}, we call the agents to be guided as sheep, and the steering agent as a shepherd. As shall be described in Subsection~\ref{subsc:sheep}, the sheep agents move on the plane according to the inter-flock dynamics and the repulsive force from the shepherd agent. The objective of the shepherd agent is set to be the guidance of \waka{Nに読める．N-Mでは？}the sheep agents \addddd{to be guided} \waka{ゴールに【収める」とは少し違うニュアンスになっているが大丈夫か}\delll{toward}\addddd{into} a goal region~$G$, which is assumed to be an open disk with center~$x_g\in \mathbb R^2$ and radius~$R_g > 0$.

% 本報告を通じて以下の記法を用いる．羊に対して番号$1$, \dots, $N$を割り振る．この番号の集合を$[N] = \{1, \dotsc, N\}$と書く．時刻$k=0, 1, 2, \dotsc$における牧羊犬の位置を$x_d(k)$, $i$番目の羊の位置を$x_i(k)$と書く．また, 集合$X$に対してその要素数を$\abs{X}$と書く．

Throughout this paper, we use the following notations. We assign the numbers~$1$, \dots, $N$ to the sheep agents. The set of these numbers is denoted as $[N] = \{1, \dotsc, N\}$. Also, we let $x_d(k)\in \mathbb R^2$ denote the position of the shepherd agent at time~$k$, and $x_i(k)\in \mathbb R^2$ denote the position of the $i$th sheep at time~$k$. For a set~$X$, we let $\abs{X}$ denote the number of elements of $X$.
% \masaki{ノルムの記法が未定義}\masaki{定義した}
For a real vector~$v$, we let $\norm{v}$ denote the Euclidean norm of~$v$.

\subsection{Sheep dynamics}\label{subsc:sheep}
% 羊の移動モデルを述べる．各時刻において$i$番目の羊の位置$x_i(k)$は, 次の式に従って更新されるものとする．
% \begin{equation}\label{eq:sheep_position}
%     x_i(k+1) = x_i(k) + v_i(k).
% \end{equation}
% ここで, $v_i(k)$は時刻$k$における羊$i$の移動量を表すベクトルである．本報告においては, ベクトル$v_i(k)$はReynolds~\cite{reynolds1987flocks}の提案したBoidモデルに基づいて
% 構成されるものとする．Boidモデルは, shepherdingモデルに基づく誘導問題の研究において広く使われているモデルである．
% 他のエージェントから離れる「分離」, 他のエージェントと速度を合わせる「整列」, 他のエージェントに近づく「結合」の三種類の相互作用がBoidモデルでは仮定される．
% この三種類の相互作用に加えて, 羊は牧羊犬を回避するように行動するものとする．

In this subsection, we present the mathematical model of the movement of sheep agents. We assume that, at each time~$k$, the position $x_i(k)$ of the $i$th sheep is updated by the difference equation
\begin{equation}\label{eq:sheep_position}
    x_i(k+1) = x_i(k) + v_i(k),
\end{equation}
where $v_i(k) \in \mathbb R^2$ denotes the movement vector of the $i$th sheep at time $k$. In this paper, we assume that the vector~$v_i(k)$ is constructed according to the Boid model~\cite{reynolds1987flocks}, a model widely used in the context of the shepherding problem~\cite{e,strombom2014solving,tsunoda2018analysis,El-Fiqi2020a,Zhi2021}
% \resolved{Shepherdingに関連する文献であってBoidモデルを使っているものを追加で2報ほど見繕ってもらえないでしょうか．}\anna{論文探し方のコツ教えていただきたいです．．}\masaki{Shepherdingやってる論文はたいていstrombomの論文を引いているとおもいます．僕ならこの論文を引用する論文\url{https://scholar.google.com/scholar?cites=1319647414415102336&as_sdt=2005&sciodt=0,5&hl=en}を片っ端からしらべてみます．追加の目的の一つは多様な人の名前がreferencesにのることです．}\anna{保留にします}\masaki{やりました．確認おねがいします．}.
In the Boid model, the following three types of inter-flock interactions are assumed: ``separative force'' from other agents, ``alignment force'' to match the speed of other agents, and ``attractive force'' to approach other agents. In addition to these three types of interactions, the sheep are assumed to move in such a way as to avoid the shepherd.
% \masaki{3/22：変更ありがとう}
% 以上を反映して\eqref{eq:sheep_position}式に現れるベクトル$v_i(k)$を次式により定める．
% \begin{equation}\label{eq:sheep_vector}
%     v_i(k) = K_{i1}v_{i1}(k) + {K_{i2}v_{i2}(k)} + K_{i3}v_{i3}(k) + K_{i4}v_{i4}(k).
% \end{equation}
% ただしここで$K_{i1}, K_{i2}, K_{i3}, K_{i4}$は個々の羊に依存する非負の定数である．また, $v_{i1}(k)$, $v_{i2}(k)$, $v_{i3}(k)$はそれぞれBoidモデルの分離, 整列, 結合に対応するベクトルである．この三種類のベクトルに, 牧羊犬からの斥力に対応するベクトル$v_{i4}$が加わっている．
Then, the vector~$v_i(k)$ appearing in equation~\eqref{eq:sheep_position} is determined as
\begin{equation}\label{eq:sheep_vector}
    v_i(k) = K_{i1}v_{i1}(k) + {K_{i2}v_{i2}(k)} + K_{i3}v_{i3}(k) + K_{i4}v_{i4}(k),
\end{equation}
where $K_{i1}$, $K_{i2}$, $K_{i3}$, and~$K_{i4}$ are non-negative constants that depend on individual sheep. Also, $v_{i1}(k)$, $v_{i2}(k)$, and~$v_{i3}(k)$ are vectors corresponding to the separative, alignment, and attractive forces
% \masaki{藤岡さん：この３つの呼び方, たぶん英訳で統一されていないと思います．確認と統一をおねがいします．}\anna{FAT論文に従って, 分離"repulsive force", 整列"alignment force", 結合"attractive force", 斥力"repulsive force from shepherd"にしました}\masaki{ありがとう了解です}\masaki{藤岡さん：やっぱり repulsion from sheep は separation でおねがい}\anna{直しました}\masaki{thanks}
of the Boid model, respectively. To these three vectors we add vector~$v_{i4}(k)$ corresponding to the repulsive force from the shepherd.

% 羊は中心$x_i(k)$, 半径$R$の円内に存在する全ての羊から力を受けるものとする．この範囲に他の羊が存在しない場合は, $v_{i1}(k) = v_{i2}(k) = v_{i3}(k) = 0$とする．時刻$k$において羊$i$から半径$R>0$以内に存在する羊の番号の集合を$S_i(k)$とすると, $v_{i1}(k)$, $v_{i2}(k)$, $v_{i3}(k)$は
% \begin{align}
%     v_{i1}(k) &= -\frac{1}{|S_i(k)|} \sum_{j \in S_i(k)} \frac{x_j(k) - x_i(k)}{\norm{x_j(k) - x_i(k)}^3}, \label{eq:sheep_v1}\\
%     v_{i2}(k) &= \frac{1}{|S_i(k)|} \sum_{j \in S_i(k)} \frac{v_j(k-1)}{\norm{v_j(k-1)}}, \label{eq:sheep_v2}\\
%     v_{i3}(k) &= \frac{1}{|S_i(k)|} \sum_{j \in S_i(k)} \frac{x_j(k) - x_i(k)}{\norm{x_j(k) - x_i(k)}} \label{eq:sheep_v3}
% \end{align}
% と定まる．
% 最後に, $v_{i4}(k)$は牧羊犬からの斥力に対応するベクトルであり,
% \begin{align}
%     v_{i4}(k) &= -\frac{x_d(k) - x_i(k)}{\norm{x_d(k) - x_i(k)}^3} \label{eq:sheep_v4}
% \end{align}
% により定まる．

We assume that the $i$th sheep receives forces from all sheep in the circle with center~$x_i(k)$ and radius~$R>0$. If there are no other sheep in this range, then we set $v_{i1}(k) = v_{i2}(k) = v_{i3}(k) = 0$. If we let $S_i(k)$ denote the set of the indices
% \resolved{意見おねがいします．「羊の添字」というのはあまり適切ではないと僕は思っています．添字は「文字の周囲に小さく添えられた文字」であって，羊は文字ではないので．むしろ羊の番号だとおもっています．番号を"index"と訳しています．indexの複数形で"indices"です．Google Scholar \url{https://scholar.google.com/scholar?q="the+set+of+the+indices"}でもそれなりの使用例があるので，このままを提案します．}\anna{ありがとうございます．``index''は「添字」という意味しかないと誤認していましたので，このままという意見に賛成です．}
of the sheep \waka{無生物に対してであればlieでいいけど羊なので寝転んでいるように思える．なくてもいいのでは．}\delll{lying }within radius~$R$ of the $i$th sheep at time $k$, then the vectors~$v_{i1}(k)$, $v_{i2}(k)$, and~$v_{i3}(k)$ are given by
\begin{align}
    v_{i1}(k) &= -\frac{1}{|S_i(k)|} \sum_{j \in S_i(k)} \frac{x_j(k) - x_i(k)}{\norm{x_j(k) - x_i(k)}^3}, \label{eq:sheep_v1}\\
    v_{i2}(k) &= \frac{1}{|S_i(k)|} \sum_{j \in S_i(k)} \frac{v_j(k-1)}{\norm{v_j(k-1)}}, \label{eq:sheep_v2}\\
    v_{i3}(k) &= \frac{1}{|S_i(k)|} \sum_{j \in S_i(k)} \frac{x_j(k) - x_i(k)}{\norm{x_j(k) - x_i(k)}} \label{eq:sheep_v3}\dell{.}\adddd{,}
\end{align}
\dell{Finally,}\adddd{whereas} $v_{i4}(k)$ \dell{ represents the vector corresponding to the repulsion from the shepherd, and }%
% \masaki{懸念の重複を避けるためにこのようにしてみました．確認おねがいします．}
is given by
\begin{align}
    v_{i4}(k) &= -\frac{x_d(k) - x_i(k)}{\norm{x_d(k) - x_i(k)}^3}.  \label{eq:sheep_v4}
\end{align}

\subsection{Guidance problem}\label{subsc:dynamics}

% 本報告では他の羊との分離, 整列, 結合, 牧羊犬からの斥力の四種類の力を全て受ける羊の中に, 四種類のうち一種類以上の力を受けないような異種の羊が$M$体混在する状況を考える．ただしここで, 同時に存在する異種の羊の種類は一つであると仮定する．異種ではない羊を通常の羊と呼ぶ．一般性を失うことなく, 通常の羊の番号を$1, 2, \dotsc, N-M$とする．異種の羊の番号は$N-M+1, N-M+2, \dotsc, N$である．

In the shepherding problem we consider in this paper, it is supposed that the sheep agents consists of the following two types of sheep; \emph{normal} and \emph{variant}. A normal sheep is assumed to be subject to all the four types of forces \eqref{eq:sheep_v1}--\eqref{eq:sheep_v4}: separation, alignment, attraction, and repulsion from the shepherd. \waka{四つの力を受けるけどパラメータが違う，というvariantの設定方法もありうるので，そうではなく力を受けないとすることの意味，目的が述べられていると良い}\masaki{端点だけ考えようね，という話ではあって，かつこの仮定を今置くひつようもなさそうで，たぶんアルゴリズム評価の段で述べるべきだった？}On the other hand, a variant sheep is assumed to be subject to at most three types of the four forces. \addddd{Therefore, we do not consider the situation in which a variant sheep receives all the four forces but with different coefficients. We place this assumption for simplicity of the formulation; in particular, the shepherding method developed in this paper is  applicable to such general cases.}\masaki{苦しい言い訳}\delll{ We assume that there exist $M$ variant sheep. Without loss of generality, the sheep $1$, $2$, \dots, $N-M$ are normal, and the sheep $N-M+1$, $N-M+2$, \dots, $N$ are variant.}

% \eqref{eq:sheep_vector}式において通常の羊がもつ係数は全て共通であり, ある正の定数$K_1$, $K_2$, $K_3$, $K_4$が存在して
% \begin{equation}\label{eq:keisu_nomal}
%     K_{i1} = K_1,\ K_{i2} = K_2,\ K_{i3} = K_3,\ K_{i4} = K_4
% \end{equation}
% が任意の$i=1, \dotsc, N-M$に対して成り立つ．異種の羊についても同様に, ある非負の実数$\alpha_{1}$, $\alpha_{2}$, $\alpha_{3}$, $\alpha_{4}$が存在して
% \begin{equation}
%     K_{i1} = \alpha_{1}K_{1},\ K_{i2} = \alpha_{2}K_{2},\ K_{i3} = \alpha_{3}K_3,\ K_{i4} = \alpha_{4}K_4
% \end{equation}
% が任意の$i=N-M+1, \dotsc, N$に対して成り立つとする．ここで, $\alpha = (\alpha_1, \alpha_2, \alpha_3, \alpha_4)$と表すものとする．本報告では簡単のため$\alpha_{i}$の取りうる値を1もしくは0に限定する．例えば, 異種の羊が分離と整列の力のみを受けるとき, $\alpha = (1, 1, 0, 0)$であるとする．

\masaki{新情報後置できなくなったのでこっちにうつした}\addddd{We assume that there exist $M$ variant sheep. Without loss of generality, we assume that the sheep $1$, $2$, \dots, $N-M$ are normal, and the sheep $N-M+1$, $N-M+2$, \dots, $N$ are variant.}
In this paper, we do not consider heterogeneity within the set of the normal sheep and the one of the variant sheep. Hence, we assume the existence of positive constants~$K_1$, $K_2$, $K_3$, and~$K_4$ such that
\begin{equation}\label{eq:keisu_nomal}
    K_{i1} = K_1,\ K_{i2} = K_2,\ K_{i3} = K_3,\ K_{i4} = K_4
\end{equation}
for all $i=1, \dotsc, N-M$. Likewise, we assume the existence of nonnegative constants~$\alpha_{1}$, $\alpha_{2}$, $\alpha_{3}$, and $\alpha_{4}$ such that
\begin{equation}
    K_{i1} = \alpha_{1}K_{1},\ K_{i2} = \alpha_{2}K_{2},\ K_{i3} = \alpha_{3}K_3,\ K_{i4} = \alpha_{4}K_4
\end{equation}
for all $i=N-M+1, \dotsc, N$. We remark that the quadruple
\begin{equation}
    \alpha = (\alpha_1, \alpha_2, \alpha_3, \alpha_4)
\end{equation}
characterizes the deviation of the dynamics of the variant sheep from that of the normal sheep.
In this paper, we suppose that \dell{the vector~$\alpha$}\adddd{each coefficient $\alpha_i$} is \dell{$\{0, 1\}$-valued}\adddd{either $0$ or $1$}. Under this assumption, for example, if a variant sheep receives only the forces of separation and alignment, then we have $\alpha = (1, 1, 0, 0)$.
% \masaki{メモ：ちょっと流れが}\masaki{$\alpha$が自然に現れるようにした}

% 以上の設定のもとで本報告における目的は, $N-M$体の通常の羊をできるだけ多く目的地$G$へ誘導するような牧羊犬の機動アルゴリズムを設計することである．\add{この誘導において牧羊犬は一定時間$T$ごとに全ての羊の位置を知ることができると仮定する．通常の羊のダイナミクスを特徴づける係数$(K_1,K_2,K_3, K_4)$の推定値$(\beta_{1}K_1, \beta_{2}K_2, \beta_{3}K_3, \beta_{4}K_4)$も利用可能であると仮定する．ここで, $\beta_i$は正の実数とする．$\beta = (\beta_{1}, \beta_{2}, \beta_{3}, \beta_{4})$とする．この推定値は正しい係数と一致しているとは限らないものとする．本報告では簡単のため, $\beta_i$の取りうる値を$0 \leq \beta_i \leq 1$に限定する．}

% \resolved{この段落では，査読者が重要な情報を見つけやすくするためにequation環境などを使って犬が取得可能な情報を以下のようにまとめてみました．正しいかどうか確認お願いします．}\anna{確認しました．多分正しいです．}\masaki{thanks}
As for the information available to the shepherd, we consider the following situation. First, we assume that the shepherd is initially given
an estimate
\begin{equation}\label{eq:estimates}
(\beta_{1}K_1, \beta_{2}K_2, \beta_{3}K_3, \beta_{4}K_4)
\end{equation}
of the coefficients $(K_1,K_2,K_3, K_4)$ characterizing the dynamics of the normal sheep. We do not require that the estimate is correct; therefore, each of the constants $\beta_1$, $\beta_2$, $\beta_3$, and $\beta_4$ is not necessarily equal to one.
% \masaki{このような設定にする理由を言う必要がありそう？}
Secondly we assume that, in the process of the guidance operation, the shepherd \waka{giveする主語は誰か？}will be given the location of all the sheep at every $T$ units of time \addddd{from an external system for observation}\masaki{苦し紛れだが}; i.e., we assume that the shepherd can obtain the set of vectors
\begin{equation}\label{eq:periodicobservation}
    \{x_i(\ell T)\}_{i\in [N]}
\end{equation}
for each $\ell \geq 0$. Finally, \addddd{in addition to this global but periodic information,}\ask{「2番目の前提との対応」を明らかにしたつもりなのですが，これでいいのでしょうか．意見ください．}\anna{時刻$T$ごとに全羊$+$毎時刻nearest sheepということですよね．}\masaki{そうです，もうこれでいいよね，これにします．}\anna{いいと思います．ありがとうございます} we suppose that the shepherd is able to measure the \dell{relative locations of a few sheep agents of its choice at each time}\adddd{position of the sheep closest to the shepherd at every time instants to avoid collision. Hence, the shepherd is assumed to know the index}
\begin{equation}\label{eq:n(k)}
    \adddd{
    n(k) = \argmin_{i\in [N]}{
    \norm{x_d(k)-x_i(k)}}.
    }
\end{equation}
\adddd{at each time $k\geq 0$.}\waka{2番目の前提との対応が分かりません．羊の正確な位置はT毎にしかわからないけど一番近い羊は常に分かると言うこと？}
% Let $\beta = (\beta_{1}, \beta_{2}, \beta_{3}, \beta_{4})$.

We can now state the objective of this paper as follows.

\begin{problem}\label{prb:}
% \masaki{査読者が問題設定を見つけやすいようにした}
Develop a movement algorithm of the shepherd so that the shepherd can  \waka{これの定義が不明}\delll{guide}\addddd{let} as many normal sheep as possible \delll{to}\addddd{arrive into} the destination area~$G$ by using the information \eqref{eq:estimates}, \eqref{eq:periodicobservation}, \eqref{eq:n(k)} given to the shepherd.
% \masaki{与えられる情報を明示した}
\end{problem}
% The objective of this paper is to

% In this guidance, we assume that For simplicity, we restrict the possible values of $\beta_i$ to $0 \leq \beta_i \leq 1$.\masaki{$\alpha$と$\beta$オッケーです}

\section{Proposed method}\label{sc:virtual_method}
% 提案する牧羊犬の誘導手法を述べる．まず\ref{subsc:gaiyou}節では，提案手法の概要を述べる．\ref{subsc:kasou_dynamics}節では提案手法において重要な役割を果たす仮想的な羊と呼ばれるエージェントのダイナミクスを述べる．この仮想的な羊を用いた異種エージェント判別に基づく牧羊犬の機動アルゴリズムを\ref{subsc:shepherd_alg}節において述べる．
% \masaki{この段落藤岡さんおねがいします}\anna{確認お願いします}\masaki{少し手直ししました}
\dell{In this section, we describe the proposed guidance method based on model predictive discrimination. In section 3.1 we describe the overall behavior of the proposed method. In section 3.2 we introduce the dynamics of the agents, called virtual sheep, which play an important role in the proposed method. Then, in section 3.3 we describe the proposed method using the virtual sheep. The proposed method is based on variant agent discrimination.}
\adddd{In this section, we describe the proposed shepherding method based on model-based discrimination of variant sheep agents. In Subsection~\ref{subsc:gaiyou}, we describe the overall behavior of the proposed method. Then, in Subsection~\ref{subsc:kasou_dynamics}, we introduce auxiliary agents called virtual sheep, which play an important role in the proposed method. A detailed description of the proposed method is presented in Subsection~\ref{subsc:shepherd_alg}.}

\subsection{Overall behavior}\label{subsc:gaiyou}
% 異種の羊が存在しない場合，つまり，全ての羊に対して\eqref{eq:keisu_nomal}式が成り立つ場合には，Farthest-Agent Targeting 法と呼ばれる手法による誘導が有効である\cite{tsunoda2019analysis}．しかしながら，異種の羊が混在する状況では，誘導が成功するとは限らない．したがって，異種の羊を除外した群れに対してFarthest-Agent Targeting 法\cite{tsunoda2019analysis}を適用することを考える．ただしここで，本報告においては，牧羊犬は異種の羊に関する情報を与えられていないと仮定した．そこで提案手法では，羊の種類を判断するために，\ref{subsc:kasou_dynamics}節および\ref{subsc:shepherd_alg}節で述べる方法を用いて予測される羊の軌道からの乖離度を用いる．
% \masaki{お化粧的に少し書き直しました}

\waka{3章のところ，位置更新と判定のタイミングの記述が問題になっているようなの
で，僕なら3.1でまず全体の流れをもう少し詳しく説明します．}
\waka{牧羊犬はノーマルだと判断している羊の群れをFATで誘導する．
ノーマルか異種かを判断するために羊の軌跡を推定する．
そのために仮想的な羊を導入する．
仮想的な羊はノーマルな羊と同じように動く．
T毎に仮想的な羊と実際の羊の位置を確認して離れていれば異種だと判定する．
一方，XXXであればそれ以前に異種と判断していた羊はノーマルに加える．
その後，すべての羊について？実際の羊の位置で仮想的な羊の位置を更新する．}\masaki{これはあきらめた．}

\add{Let us first describe the overall behavior of the proposed method.} We first remark that, in the special case where a variant sheep does not exist, that is, when all the sheep are normal, then applying the FAT method~\cite{tsunoda2019analysis} can be considered to be effective to solve Problem~\ref{prb:}. However, as shown by the authors in~\cite{himo2022iterative}, existence of a variant sheep would prevent the successful guidance by the FAT method. \waka{
なお，3.1の6行目は例えば"only to the flock of normal sheep in all sheep"
のように異種が混じっている中でnormalだけにFATを適用するという状況を考え
ていることを明確にした方がいいと思います．そうでないとnormalだけからなる
ホモな群れにFATを適用するのがone possibilityだと言っているようにも読める．}
One possibility in this context is applying the FAT method only to the flock of normal sheep \addddd{in all sheep}\textbf{\masaki{書き直しました，原文では正しかったので，すみません}}. However, in this paper, we are assuming that the shepherd is not given the labels (i.e., normal or variant) of sheep. In order to overcome this limitation, in this paper we propose that the shepherd performs discrimination of sheep agents by using their degree of deviation from their predicted trajectory. The details of the prediction method is presented in Subsection~\ref{subsc:kasou_dynamics}, and that of the discrimination method is presented in Subsection~\ref{subsc:shepherd_alg}.

% 提案手法では，通常の羊のダイナミクスを特徴づける係数の推定値$\beta_{i}K_i$に基づいてそれぞれの羊の軌道を予測する．係数の推定値$\beta_{i}K_i$がその実際の値$K_i$と近く，かつ異種の羊の係数$\alpha_{i}K_i$と十分に離れている場合には，通常の羊に対する軌道の予測誤差は異種の羊に対する予測誤差よりも小さいことが想定される．この仮説に基づき提案手法では，予測誤差が大きな羊は異種の羊であると判断し，牧羊犬による誘導の対象から外す．牧羊犬は，この判断に基づいて誘導の対象と定める羊のみをFarthest-Agent Targeting 法により誘導する．

For prediction of the sheep's trajectory, the proposed method uses the coefficients~$(\beta_{1}K_1, \beta_{2}K_2, \beta_{3}K_3, \beta_{4}K_4)$ given to the shepherd as an estimate of the coefficient characterizing the normal sheep. If the estimate is accurate, i.e., if the estimate~$(\beta_{1}K_1, \beta_{2}K_2, \beta_{3}K_3, \beta_{4}K_4)$ is closer to the normal coefficients $(K_1, K_2, K_3, K_4)$ than to the variant coefficients~$(\alpha_1 K_1, \alpha_2 K_2,  \alpha_3 K_3,  \alpha_4 K_4)$, then we can expect that the trajectory prediction for normal sheep is \dell{better}\adddd{more accurate}
% \resolved{修正しました，確認おねがいします}\anna{ありがとうございます}
than  that for variant sheep. Based on this supposition, the proposed method determines that sheep with larger prediction errors are  variant  and, then, excludes them from the shepherd's navigation. Specifically, the shepherd uses the FAT method to guide only those sheep \adddd{not} discriminated to be \dell{normal}\adddd{variant}.

\subsection{Dynamics of virtual sheep}\label{subsc:kasou_dynamics}
% 羊の軌道を予測するために仮想的な羊と呼ばれるエージェントを導入する．このエージェントは実際の羊それぞれに対応して一体ずつ存在するものとする．$N$体の仮想的な羊は実際の羊と同様に二次元平面$\mathbb R^2$上を運動する．時刻$k$における$i$番目の仮想的な羊の位置を$\xi_i(k)$と書く．
% \masaki{この段落藤岡さんおねがいします}\anna{確認お願いします}\masaki{ありがとう少し手直ししました}
In this subsection, we formally introduce the agents called virtual sheep\adddd{, which we use to perform the trajectory prediction of the actual sheep}.
These agents are assumed to be placed one for each of the actual sheep agents.
These $N$ virtual sheep move on \adddd{the field} $\mathbb R^2$ in \dell{the same way as}\adddd{a way similar to that of} the actual sheep\adddd{, baesd on the estimated coefficients~$(\beta_{1}K_1, \beta_{2}K_2, \beta_{3}K_3, \beta_{4}K_4)$ given to the shepherd}.
% These agents are assumed to exist one \masaki{文法？} for each of the actual sheep.\masaki{新情報の後置の原則に反している}$\xi_i(k)$ denotes the position of the $i$th virtual sheep at step $k$. \masaki{よろしく！}
\dell{The position of the $i$th virtual sheep at step $k$ is denoted by $\xi_i(k)$.}

\adddd{Let the position of the $i$th virtual sheep at time $k$ be  denoted by $\xi_i(k)$. Then, }the change in position of the virtual sheep is specified as
% \begin{equation}
% \label{eq:vsheep_position}
% \xi_i(k+1) = \left\{ \begin{array}{l}
% \displaystyle x_i(k) \;\;\;\;\;\;\;\;\;\;\;\;\;\;\;:\mathrm{if} \; k \bmod T = 0 \\
% \xi_i(k) + \phi_i(k) \;\;\; :\mathrm{otherwise}
% \end{array} \right.
% \end{equation}
\begin{equation}
\label{eq:vsheep_position}
\xi_i(k+1) =
\begin{cases}
\displaystyle
% \dell{
x_i(k)
% }
% \adddd{x_i(k)+\phi_i(k)}
, &\mbox{if $k \bmod T = 0$},
\\
\xi_i(k) + \phi_i(k), &\mbox{otherwise.}
\end{cases}
\end{equation}
\resolved{計算してから重ね直しの件，上のように式を変えれば良いとおもいます．距離計算してから重ね直し．．．はちょっとスマートじゃない感があるので．対応して下の記述は修正必要．とりあえず上の式でも大丈夫かどうか確認お願いします．}
\anna{式確認しました．$T, \dotsc$で距離計算，$T+1, \dotsc$で実羊と仮想羊が重なる，と解釈しましたが正しいでしょうか}\masaki{上だと実羊と仮想羊は永遠に重ならないですね．というか式としても正しくないですね．もとに戻しました．下に書いたようにこのまま投稿することを提案しますが，藤岡さん必要と判断するばあいは原稿の修正おねがいします．}
% \masaki{藤岡さん：cases環境を使えばすっきり書けますし，すっきりしていたほうが出版社も楽です．他もcases環境つかって書いておいてください．}\anna{直しました}\masaki{thanks}
In this equation, we assume that the position of the $i$th virtual sheep is re-positioned to the same position as the (actual) $i$th sheep at every $T$ units of time, when a global measurement~\eqref{eq:periodicobservation} of the sheep positions become available to the shepherd. \ask{重ね直しの理由を削除した（してしまった）理由を思い出しました．globalな観測をする度に重ね直すのは普通のように思えます．なので，重ね直す理由をことさら述べるのには違和感がありました．仮に重ね直さないのであれば，理由は述べるべきだと思います．この理由であまり筆が進みませんでした．なので折衷案として重ね直しの「効果」を述べるようにしてみました．いかがでしょうか．}\anna{ありがとうございます}
\addddd{This re-positioning allows us to prevent an unlimited growth of the distance between the normal and the virtual sheep caused by their difference in dynamics, which is therefore necessary for performing discrimination effectively.}
%
% \masaki{新情報後置できなくなったので改段落}
Furthermore, $\phi_i(k)$ is the vector representing the movement of the virtual sheep at time~$k$, and is determined in a way similar to equation~\eqref{eq:sheep_vector} for the actual sheep as
\begin{equation}\label{eq:vsheep_vector}
    \phi_i(k) = \beta_{1}K_{1}\phi_{i1}(k) + \beta_{2}K_{2}\phi_{i2}(k) + \beta_{3}K_{3}\phi_{i3}(k) + \beta_{4}K_{4}\phi_{i4}(k),
\end{equation}
where $\beta_{1}K_1$, $\beta_{2}K_2$, $\beta_{3}K_3$, and $\beta_{4}K_4$ are the estimated coefficients of the normal sheep and are given in advance to the shepherd. Also, $\phi_{i1}(k)$, $\phi_{i2}(k)$, $\phi_{i3}(k)$, and~$\phi_{i4}(k)$ are vectors corresponding to separation, alignment, attraction, and repulsion from the shepherd, respectively. Because the global measurement \eqref{eq:periodicobservation} is available only periodically, between two consecutive global measurements, the virtual sheep is assumed to perform its motion with reference to the virtual sheep's position and displacement. Therefore, the vectors in the equation~\eqref{eq:vsheep_vector} are constructed as
\begin{align}
    \phi_{i1}(k) &= -\frac{1}{|T_i(k)|} \sum_{j \in T_i(k)} \frac{\xi_j(k) - \xi_i(k)}{\|\xi_j(k) - \xi_i(k)\|^3}, \label{eq:vsheep_v1:virtual}\\
    \phi_{i2}(k) &= \frac{1}{|T_i(k)|} \sum_{j \in T_i(k)} \frac{\phi_j(k-1)}{\|\phi_j(k-1)\|}, \label{eq:vsheep_v2:virtual}\\
    \phi_{i3}(k) &= \frac{1}{|T_i(k)|} \sum_{j \in T_i(k)} \frac{\xi_j(k) - \xi_i(k)}{\|\xi_j(k) - \xi_i(k)\|}, \label{eq:vsheep_v3:virtual}\\
    \phi_{i4}(k) &= -\frac{x_d(k) - \xi_i(k)}{\|x_d(k) - \xi_i(k)\|^3},  \label{eq:vsheep_v4:virtual}
\end{align}
where $T_i(k)\subset [N]$ is the set of \dell{numbers}\adddd{indices} of the virtual sheep within radius $R$ of the $i$th virtual sheep at time~$k$, and is defined by
\begin{equation}\label{eq:bangou:virtual}
    T_i(k) = \{j\in [N]\backslash \{i\}
    \mid
    \norm{\xi_j(k) - \xi_i(k)}\leq R
    \}.
\end{equation}
If $T_i(k)$ equals the empty set, then we set~$\phi_{i1}(k) = \phi_{i2}(k) = \phi_{i3}(k) = 0$.

\subsection{Movement algorithm of shepherd}\label{subsc:shepherd_alg}

% 提案するアルゴリズムはFarthest-Agent Targeting法に基づく．Farthest-Agent Targeting法ではtarget sheepと呼ばれる羊が全ての羊の中から選ばれるのに対して，提案アルゴリズムでは牧羊犬により異種の羊でないと判断された羊の集合$I \subset [N]$から選ばれる．\add{また，Farthest-Agent Targeting法では牧羊犬は毎時刻任意の羊の位置を観測できると仮定している．しかし，提案アルゴリズムでは牧羊犬は一定時刻$T$ごとにのみ任意の羊の位置を観測できるものとする．これ以外の時間は，target sheepの位置と，牧羊犬に最も近い羊の位置のみを取得できるものとする．}

We are now ready to describe the proposed movement algorithm of the shepherd.
The proposed \dell{algorithm}\adddd{method} is based on the FAT method. In the original FAT method~\cite{tsunoda2019analysis}, a sheep called the target sheep is selected from among all the sheep. On the other hand, the proposed \dell{algorithm}\adddd{method} selects sheep only from those discriminated to be normal by the shepherd. The discrimination is performed by using the virtual sheep introduced in Subsection~\ref{subsc:kasou_dynamics}.
% The Farthest-Agent Targeting method assumes that the shepherd can observe the position of any sheep at any given time.
% \masaki{要確認：以下はSubsection~\ref{subsc:dynamics}と重複？}\masaki{思い出させてあげる感じにすることにした．}However, in our problem setting, we assume that the shepherd can observe the position of all the sheep only periodically. At other times, only the position of the target sheep and the position of the sheep closest to the shepherd can be obtained.

% 集合$I$は初期時刻において$I=[N]$と初期化され，一定時間$T$ごとに以下のように更新される．牧羊犬はまず，全ての羊に対して，後述の手法により羊の種類を判断し，異種と判断された羊$i$を集合$I$から取り除く．次に，集合$I$から取り除かれた羊全てに対して，後述する一定の条件を満たした羊を再び集合$I$に加える．

In the algorithm, the shepherd possesses as its internal variable a set~$I(k) \subset [N]$, which is used to record the set of the sheep index discriminated to be normal. The shepherd first initializes this set as $I(0)=[N]$ and update the set with period~$T$. At each time instant for update, a sheep index that is discriminated to be variant is removed from the set. The discrimination is performed by the rule described later in this subsection. On the other hand, for all the sheep that have been once removed from the set~$I(k)$, the shepherd checks a condition described below. If the sheep satisfies the condition, then its corresponding index is recovered into the set.

% まず，羊の種類を判断する手法を述べる．牧羊犬は，全ての羊$i\in  [N]$について，その位置$x_i(k)$とその羊に対応する仮想的な羊の位置$\xi_i(k)$との間の距離$\norm{x_i(k) - \xi_i(k)}$を求める．この距離が後述する方法により定めた閾値よりも大きい場合は，羊$i$を異種の羊であると判定する．この時，$i$が集合$I$に含まれている場合にはこの集合から取り除く．
For discrimination, with period~$T$, the shepherd computes the distance $\norm{x_i(k)- \xi_i(k)}$ \resolved{この距離を計算するのは時間が$T-1$, $2T-1$, $3T-1$, \dots のときですか？というのもk mod T = 0 のときは重ね直されるので距離が0になってしまうので．仮にそうだとすると，そこは説明必要かなとおもいます．}\anna{プログラム上はTの時に距離計算です．計算してから重ね直しています．距離計算することを先に説明すべきですか？}\masaki{了解です．想像どおりでした．でしたら式\eqref{eq:vsheep_position}を変更することを提案します．式\eqref{eq:vsheep_position}を変えてみましたので確認してください．}\resolved{上記について（式とアルゴリズムを整合させられるような）適当な対応策を思いつかなくて，かつ計算してから重ね直しをするのはある意味自明なので，
現状の記述で投稿することを提案します．もし藤岡さんが気になるようであれば，必要な修正を日本語でも構わないのでどこかにいれておいてください．必要あれば英訳は小蔵がやります．}\anna{このままにします} between the actual and virtual sheep for all  $i\in [N]$. If this distance is greater than a threshold value \adddd{$L$}, then the $i$th sheep is discriminated to be variant and, therefore, is removed from the set~$I(k)$.
% \masaki{要確認：ここらへん説明が後ろ回しになっているのが入れ子になっているので構造を変えること}\masaki{心配事は一つにします．}
% 本報告では距離の閾値の定め方を二つ提案する．一つ目の定め方は，固定した閾値を用いる手法である．二つ目の定め方は，四分位範囲を用いて閾値を動的に定める手法である．四分位範囲を用いた閾値の設定方法は，統計処理における外れ値検出の方法として広く使われている
%\cite{hazureti}．
% 上述した実際の羊の位置と仮想的な羊の位置との距離$\norm{x_i(k) - \xi_i(k)}$に対して，第一四分位数$q_{1}$および第三四分位範数$q_{3}$を用いて四分位範囲$ \mathrm{IQR} = q_{3} - q_{1}$を求める．距離の閾値$L$を
% \begin{equation}\label{eq:L_dynamic}
%     L =  q_{3} + 1.5\;\mathrm{IQR}
% \end{equation}
% と定める．
% ここで，一つ目の定め方をStatic手法，二つ目の定め方をDynamic手法と呼ぶ．
As for determining the threshold, we propose the following two methods; Static and Dynamic. The Static method uses a fixed threshold, while the Dynamic method dynamically and adaptatively sets the threshold using quartile ranges, a common outlier detection method in statistics. Specifically, in the latter approach, the first quartile~$q_{1}$ and third quartile~$q_{3}$ are first computed for the set~$\{\norm{x_i(k) - \xi_i(k)}\}_{i \in [N]}$ of the distances. Then, the distance threshold $L$ is determined as
\begin{equation}\label{eq:L_dynamic}
    L = q_{3} + 1.5\;\mathrm{IQR}
\end{equation}
with the interquartile range~$\mathrm{IQR} = q_{3} - q_{1}$.
% The first method is called the Static method and the second method is called the Dynamic method.

% 牧羊犬は次に，異種と判定されている全ての羊$i \notin I$に対して，その羊が異種であると判定された回数が閾値$n$回以下であり，かつ最後に異種であると判定されてから時間$h$経過している場合には，番号$i$を集合$I$に加える．したがって，ある羊の番号が集合$I$から恒久的に取り除かれるためには，その羊が$n+1$回異種であると判定される必要がある．この仕組みを設ける理由は，係数の推定値$\beta_{i}K_i$における誤差により上述の判定法が正確に行われるとは限らないためである．
% この不正確性を補うことを意図して，本報告では判定アルゴリズムに閾値$n$を設けている．

% \masaki{この文がわかりづらい}
At each update time of the set~$I(k)$, once the shepherd finishes discrimination of all the sheep agents, the shepherd then adds the index~$i$ to the set~$I(k)$ for all sheep~$i \notin I(k)$ that have been determined to be variant, if the number of times the sheep has been determined to be variant is less than the pre-determined threshold $\tau>0$
% \resolved{この$n$は違う記号の必要があります．というのも$n$は既に最近個体に使われているので．適切な記号を検討して，かつ変更してください．}\anna{a or m と考えました．今はaにしています．適切でなければ教えていただけますでしょうか}\masaki{ありがとう．ふつうのローマンだとちょっと埋もれてしまうので，ギリシャの$\tau$にしました．threshold の頭文字のギリシャ文字です．論文中で他にこの記法が現れる部分があれば更新お願いします．}
and, furthermore, $h$ units of time has passed since its last discrimination to be variant. Thus, for a sheep to be permanently removed from the set~$I(k)$, the sheep must be determined to be variant $\tau+1$ times. The reason for introducing this mechanism is that the above decision method is not always accurate due to the potential error in the coefficient estimate  given to the shepherd.  \textbf{\anna{日本語(説明不足な気がしています) : また，variant sheep の近くに存在するsheepは，その種類に関わらず，actual sheepとvirtual sheepの軌道がずれてしまう可能性がある．variant sheepと，それに対応するvirtual sheepの位置は異なるが，周囲のsheepは，actual sheep同士，virtual sheep同士で力を受けるためである．}}{\ask{
こういうこと．．．でしょうか．確認おねがいします．30分かけました（他意はないです）．}}\anna{ありがとうございます．こういうことを言いたかったです...}\addddd{Even when the estimate is correct, the non-vanishing distance between a variant sheep and its corresponding virtual sheep causes the difference in the trajectories of a normal sheep and its corresponding virtual sheep, particularly when the normal sheep has relatively many variant neighbors.} With the intention of compensating for this imprecision, we set the threshold~$\tau$ in the proposed decision algorithm.

% 牧羊犬はこのようにして構成された集合$I$に対して，文献\cite{tsunoda2019analysis}で提案されたFarthest-Agent Targeting法に基づき以下のように誘導を行う．まず時刻$k$から$k+1$における牧羊犬の位置の変化は
% \begin{equation}
% \label{eq:shepherd_position}
%     x_d(k+1) = x_d(k) + v_d(k)
% \end{equation}
% と表される．ここで$v_d(k)$は時刻$k$における牧羊犬の移動量を表すベクトルであり，
% \begin{equation}
% \label{eq:shepherd_vector}
%     v_d(k) = K_{d_1}v_1(k) + K_{d_2}v_2(k) + K_{d_3}v_3(k)
% \end{equation}
% により定まる．この式において$K_{d_1}$, $K_{d_2}$, $K_{d_3}$は正の定数であり，またベクトル$v_1(k)$, $v_2(k)$, $v_3(k)$は
% \begin{align}
%     v_1(k) &= \frac{x_{d}(k) - x_{t(k)}(k)}{\|x_d(k) - x_{t(k)}(k)\|}, \label{eq:shepherd_v1}\\
%     v_2(k) &= -\frac{x_d(k) - x_{n(k)}(k)}{\|x_d(k) - x_{n(k)}(k)\|^3}, \label{eq:shepherd_v2}\\
%     v_3(k) &= -\frac{x_d(k) - x_g}{\|x_d(k) - x_g\|} \label{eq:shepherd_v3}
% \end{align}
% により定まる．$t(k)$はFarthest-Agent Targeting法におけるtarget sheepに相当する羊の番号である．ただしここで，牧羊犬は一定時間$T$ごとにのみ実際の羊の位置を観測でき，それ以外の時間はtarget sheepと牧羊犬に最も近い羊の位置のみを観測できると仮定しているため，次のように更新される．
% \begin{equation}\label{eq:target}
%     t(k) = \left\{ \begin{array}{l}
%     \displaystyle  \argmax_{i \in I} \norm{x_i(k)-x_g}\;\;:\mathrm{if} \; k \bmod T = 0 \\
%     t(k-1) \;\;\;\;\;\;\;\;\;\;\;\;\;\;\;\;\;\;\;\;\;:\mathrm{otherwise}
%     \end{array} \right.
% \end{equation}
% また，$n(k) = \argmin_{i\in N} \norm{x_d(k)-x_i(k)}$
% は全ての羊の中で最も牧羊犬に最も近い羊の番号である．

For the set~$I(k)$ thus constructed, the shepherd performs guidance based on the FAT method proposed in~\cite{tsunoda2019analysis} as follows. First, we let the shepherd change its position from time~$k$ to~$k+1$ as
\begin{equation}
\label{eq:shepherd_position}
    x_d(k+1) = x_d(k) + v_d(k),
\end{equation}
where $v_d(k)$ is the vector of shepherd movements at time~$k$ and is constructed as
\begin{equation}
\label{eq:shepherd_vector}
    v_d(k) = K_{d_1}v_1(k) + K_{d_2}v_2(k) + K_{d_3}v_3(k).
\end{equation}
In this equation, $K_{d_1}$, $K_{d_2}$, and~$K_{d_3}$ are positive constants, and the vectors $v_1(k)$, $v_2(k)$, and~$v_3(k)$ are given by
\begin{align}
    v_1(k) &= \frac{x_{d}(k) - x_{t(k)}(k)}{\|x_d(k) - x_{t(k)}(k)\|}, \label{eq:shepherd_v1}\\
    v_2(k) &= -\frac{x_d(k) - x_{n(k)}(k)}{\|x_d(k) - x_{n(k)}(k)\|^3}, \label{eq:shepherd_v2}\\
    v_3(k) &= -\frac{x_d(k) - x_g}{\|x_d(k) - x_g\|},  \label{eq:shepherd_v3}
\end{align}
where $t(k)$ denotes the estimated index of the sheep farthest from the goal among those discriminated to be normal, and is constructed as
\begin{equation}
\label{eq:target}
t(k) =
\begin{cases}
\argmax_{i \in I(k)}{\norm{x_i(k)-x_g}}, &\mbox{if $k \bmod T = 0$},
\\
t(k-1), &\mbox{otherwise,}
\end{cases}
\end{equation}
while $n(k)$ is the index  of the sheep closest to the shepherd and is defined as in \eqref{eq:n(k)}.
We remark that, in the original FAT method, the shepherd is designed to steer the sheep farthest from the goal among all the sheep agents. However, since we assume that the shepherd can observe the actual positions of all the sheep only periodically, we alternatively adopt the formula~\eqref{eq:target} as the estimate of the normal sheep farthest from the goal.
% 異種の羊でないと判断された羊の集合$I$に含まれる羊が全て目的地$G$に属した時アルゴリズムは終了する．上述のアルゴリズムのフローチャートをFigure~\ref{fig:flowchart}に示す．

\begin{figure}[tb]
    \centering
    \includegraphics[width=.7\linewidth]{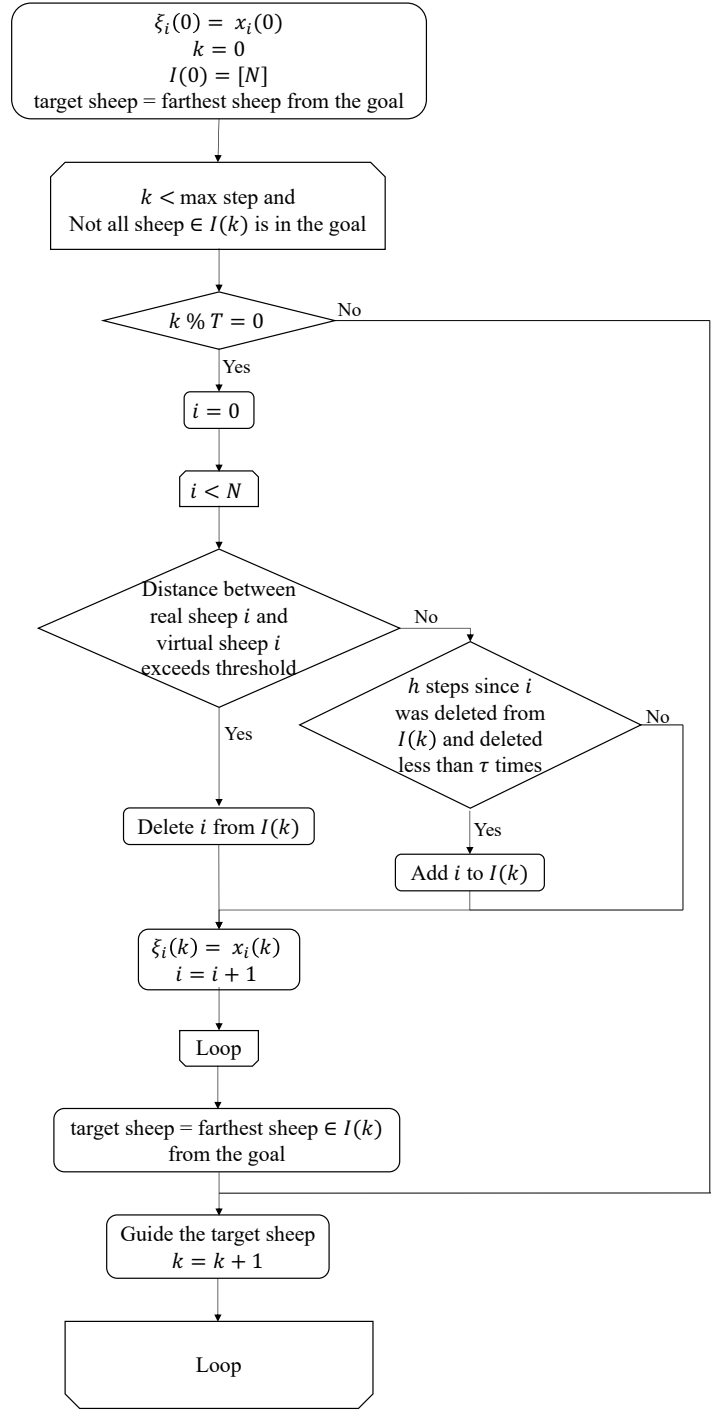}
    % \caption{提案手法のフローチャート}
    \caption{Flowchart of the proposed shepherding method
    % \anna{iPad変更点メモを反映}\masaki{thanks}
    }
    \label{fig:flowchart}
\end{figure}

Finally, the proposed \dell{algorithm}\adddd{method} terminates when all the sheep in the set~$I(k)$ are in the destination area~$G$. A flowchart of the entire algorithm presented in this section is shown in Figure~\ref{fig:flowchart}.

\section{Numerical simulations}\label{sc:sml_virtual}
% シミュレーションにより提案手法の有効性を示す．アルゴリズムの性能指標として，通常の羊のうちアルゴリズム終了時に目的地まで誘導できた羊の数を用いる．まず\ref{subsc:sml_setting}節では，このシミュレーションにおける設定を述べる．次に\ref{subsc:result}節では，提案手法とFarthest-Agent Targeting 法の比較，および提案手法間の比較を行う．

% \masaki{この段落おねがいします}\anna{確認お願いします}\masaki{Thanks少し手直ししました}
\addd{The objective of this section is to demonstrate} the effectiveness of the proposed method \addd{through numerical simulations}. As \del{a}\addd{the} performance measure \del{of the algorithm}\addd{for the comparison of  shepherding methods}, we \del{use}\addd{adopt the guidance success rate, which is defined as} the number of \addd{the normal} sheep that could be guided to the destination area at the end of executing the algorithms\del{ out of the normal sheep}. \addd{In} Subsection~\ref{subsc:sml_setting}\add{, we} first \del{describes}\addd{describe} the setup \del{for this simulation}\addd{of our simulations}. \addd{We then,}\del{Next} \addd{in} Subsection~\ref{subsc:result}\addd{, present our comparison of}\del{compares} the proposed method with the FAT method and, furthermore, \addd{the comparison} among the proposed methods (i.e., Static and Dynamic).

\subsection{Simulation setting}\label{subsc:sml_setting}
% \ref{subsc:sheep}節で述べた羊の移動モデルにおけるパラメータはTable~\ref{tbl:plm_sheep}のように設定した．また，\ref{subsc:shepherd_alg}節で述べた牧羊犬の移動モデルにおけるパラメータはTable~\ref{tbl:plm_shepherd}のように設定した．これらのパラメータは，不均質なエージェント群に対する誘導手法を提案したHimoら\cite{himo2022iterative}の手法で使用されていた値である．

\begin{table}[tb]
\tbl{Coefficients of normal sheep dynamics~\eqref{eq:sheep_vector}}
{\begin{tabular}{ccc}
Notation & Description & Value \\
\hline
$K_1$ & Separation & $100$ \\
$K_2$ & Alignment & $0.5$ \\
$K_3$ & Attraction & $2$ \\
$K_4$ & Repulsion & $500$ \\
$R$ & Radius of recognition range & $20$ \\
\end{tabular}}
\label{tbl:plm_sheep}
% \end{table}
\vspace{3mm}
% \begin{table}[tb]
\tbl{Coefficients of shepherd dynamics \eqref{eq:shepherd_vector}}
{\begin{tabular}{ccc}
Notation & Description & Value \\
\hline
$K_{d_1}$ & Attraction to $t(k)$th sheep & $10$ \\
$K_{d_2}$ & Separation from $n(k)$th sheep  & $200$ \\
$K_{d_3}$ & Repulsion from goal & $4$ \\
\end{tabular}}
\label{tbl:plm_shepherd}
\end{table}

% 全てのシミュレーションを通じて羊の総数$N$は20，シミュレーションの最大ステップを10000とした．初期時刻においてそれぞれの羊は中心(0, 0)，半径60の円の内部に一様分布に従いランダムに配置されるものとした．牧羊犬の初期位置を$(-30, -50)$，目的地の位置は中心$x_g$を(20, 20)，半径$R_g$を15とした．牧羊犬が任意の羊の位置を観測できる周期$T$を10とした．あらかじめ定めた閾値を用いるStatic手法においては，閾値の標準値を5と設定した．\add{また，モデル化誤差の大きさは簡単のため，以下のように決定するものとした．\eqref{eq:vsheep_vector}式における仮想的な羊のベクトルの係数$\beta_{i}K_i$は，異種の羊が受ける力については$\beta_{i} = 1$，異種の羊が受けない力については$\beta_{i} = 0.9$であるとした．例えば，異種の羊が分離と結合の力のみを受けるとき，つまり$\alpha = (1, 0, 1, 0)$のとき，$\beta = (0.9, 1, 0.9, 1)$である．}
Throughout our numerical simulations, the coefficients in the dynamic model of the normal sheep (equations~\eqref{eq:sheep_vector} and~\eqref{eq:keisu_nomal}) and those of the shepherd~\eqref{eq:shepherd_vector} are set as in Tables~\ref{tbl:plm_sheep} and~\ref{tbl:plm_shepherd}, respectively. These values of the coefficients are the same as the ones used in~\cite{himo2022iterative}.
We set the total number~$N$ of the sheep to be $20$, and the maximum simulation step to be $10000$ throughout all simulations. We assumed that, at the initial time~$t=0$, each sheep is randomly placed according to a uniform distribution on the open disk with center at the origin and radius~$60$. The initial position of the shepherd is set to be $(-30, -50)$. On the other hand, as for the destination area, we set its center~$x_g$ to be $(20, 20)$ and its radius to be $15$. We set the period $T$ at which a shepherd can observe the position of arbitrary sheep to $10$. \textbftemp{\addddd{We have also set the time interval $h$ for re-including a sheep  once discriminated to be variant into the set $I(k)$ to be $20$, and the threshold $\tau$ for permanently removing a sheep from the target of guidance to be $5$.}}{\resolved{「tauとhがstatic dynamic共通なのだが，そのように見えない問題がある」に対処しました．以下も同様です．}}\anna{ありがとうございます}
\dell{Also, we set is threshold to be $5$. We have also set its other parameters $\tau$ and $h$ to be $5$ and $20$, respectively.} \addddd{Finally, we set the distance threshold in the Static method as $L=5$}.
% \resolved{$\tau$と$h$の値の記述が無いような気がします．もしそうだったら追記してください．}\anna{``$\tau$の値を5，$h$を20と設定した．''を追記しました．}\masaki{thanks微調整しました．確認してください．}\anna{ありがとうございます}

As for the coefficients~$\beta_1$, $\beta_2$, $\beta_3$, and $\beta_4$ appearing in the estimates~\eqref{eq:estimates} given to the shepherd, we set $\beta_{i} = 1$ for the forces received by the variant sheep, and set $\beta_{i} = 0.9$ for the forces not received by a variant sheep. For example, when a variant sheep is subject to only separative and attractive forces, i.e., if $\alpha = (1, 0, 1, 0)$, then we set $\beta = (1, 0.9, 1, 0.9)$.
% \masaki{ここでvariant typeが14であることを述べる必要．}
\adddd{In our simulations, we consider the variance types characterized by the vectors $(\alpha_1, \alpha_2, \alpha_3, \alpha_4)$ ($\alpha_i \in \{0, 1\}$, $i=1, 2, 3, 4$) except for the trivial cases of $(1, 1, 1, 1)$ and $(0, 0, 0, 0)$. Hence, there are the total of $14$ types of variant sheep considered in our simulations.}

% 上述の設定を用いて，羊の初期配置を100通りランダムに生成し，シミュレーションを行った．この100回のシミュレーションにおいて，アルゴリズム終了時に目的地に存在する通常の羊の個体数の平均値を，誘導率と呼ぶ．
% \masaki{英訳おねがいします}\anna{確認お願いします}\masaki{OKです．微調整しました．}
Using the settings described above, we randomly generated 100 initial arrangements of sheep and, then, performed simulations. We define the guidance success rate as the average number of normal sheep that could be guided to the destination at the end of the algorithm in these $100$ simulations.

\subsection{Simulatoin results}\label{subsc:result}

% 提案手法とFarthest-Agent Targeting 法の比較および提案手法間の比較を行う．\ref{subsubsc:fat-pro}節では，異種の羊の個体数M を1 から羊の総数の半分である10 までのそれぞれの場合についてシミュレーションを行い，従来手法と提案手法を比較する．
% \ref{subsubsc:pro-dousi}節では，二つの提案手法を比較する．通常の羊を誤って異種と判断した回数，閾値の大きさと，誘導率の関係を比較する．
In this subsection, we perform comparison of the proposed methods and FAT method, as well as the comparison among the proposed methods. Specifically, in Subsection~\ref{subsubsc:fat-pro}, we compare the performance of the proposed and conventional methods by observing the dependency of their guidance success rate on the number of variant sheep. Then, in Subsection~\ref{subsubsc:pro-dousi}, we compare the two proposed methods thorough the evaluation of their frequency of misjudgement, i.e., the number of times at which a normal sheep is erroneously discriminated as variant.

\subsubsection{Comparison of the proposed and conventional methods}\label{subsubsc:fat-pro}
% 従来手法と提案手法の誘導率を比較する．Farthest-Agent Targeting 法は特定の羊をtarget sheepとして誘導する．しかしながら，target sheepが異種の羊であった場合には誘導が失敗することがある．Farthest-Agent Targeting 法による誘導失敗の例をFigure~\ref{fig:fat_failure}に示す．それに対して提案手法では，異種の羊が存在する状況においても通常の羊を誘導できる．提案手法の動作の様子をFigure~\ref{fig:proposed_success}に示す．異種の羊を検出することにより，通常の羊の誘導に成功した．

% We first illustrate the behavior of the proposed and conventinal methods.
% We compare the guidance rates of conventional and proposed methods. An example of such a failed guidance by the Farthest-Agent Targeting method is shown in Figure~\ref{fig:fat_failure}. In contrast, the proposed method can guide normal sheep even in the presence of sheep of different species. The operation of the proposed method is shown in Figure~\ref{fig:proposed_success}. It can be seen that the detection of a sheep of a different species allows successful guidance of a normal sheep.

\begin{figure}[tb]
\centering
\subfloat[\adddd{FAT method.} The guidance fails because the sheepdog keep trying to guide a variant sheep.
% \anna{確認お願いします}\masaki{概ねOKです}
\label{fig:fat_failure}]{%
      \includegraphics[width=.9\linewidth]{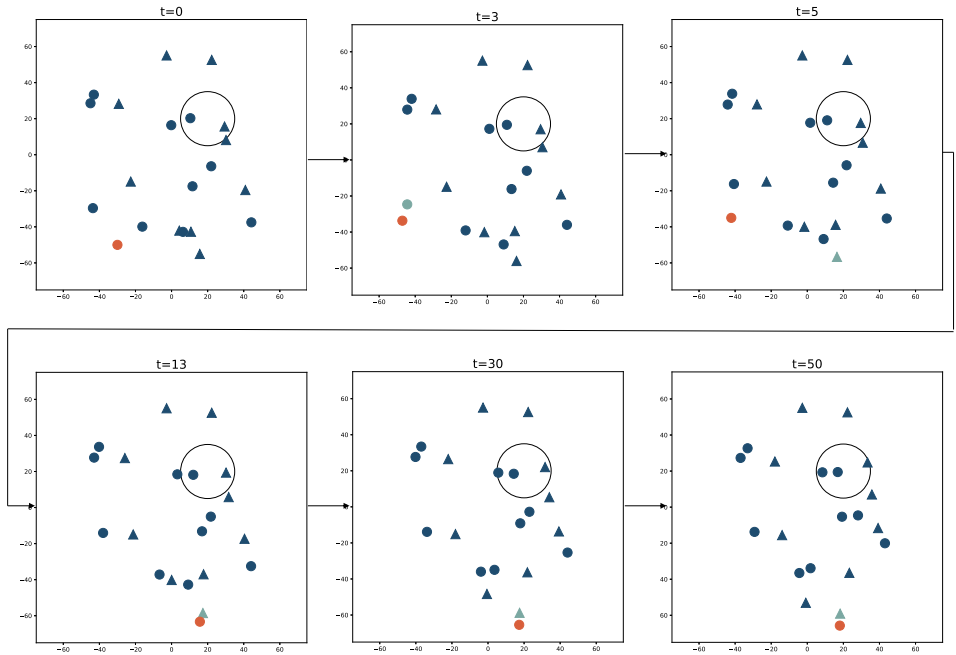}
    }
    \hfill
    \subfloat[{Static method.} The shepherd guide only normal sheep by not guiding the variant sheep.\label{fig:proposed_success}]{%
      \includegraphics[width=.9\linewidth]{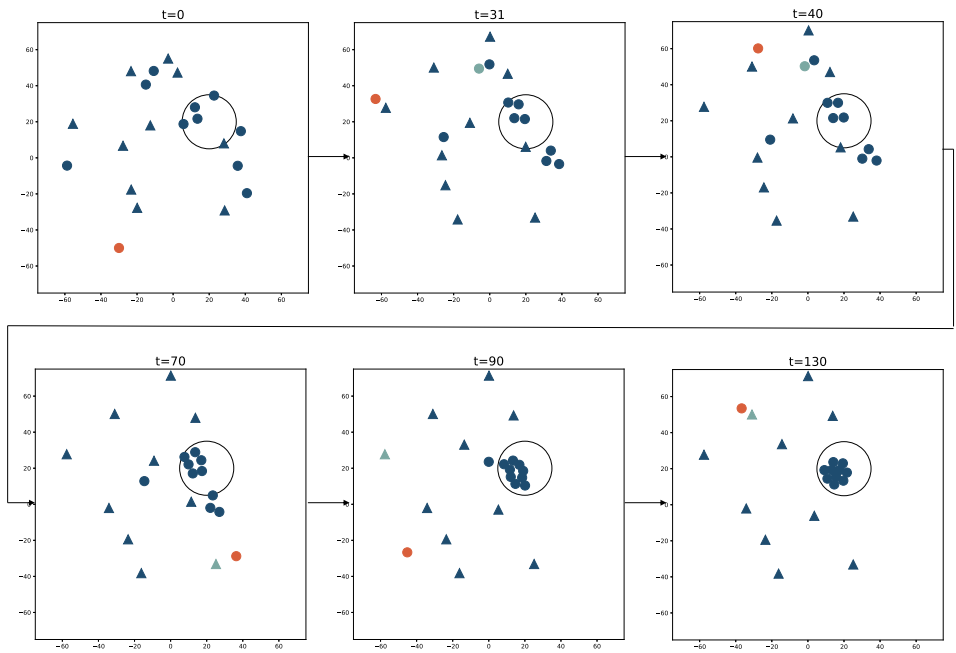}
    }
    \caption{Timeline of the guidance by the FAT \adddd{and the Static} method. The red circle indicate the sheepdog, the circles indicate normal sheep, and the triangles indicate variant sheep receive only the force of separation. The light blue represents the target sheep at the time. }
    % \label{fig:dummy}
\label{animals}
\end{figure}

% \masaki{Deepl結果を大きく変更}
We first demonstrate the overall behaviors of the proposed and conventional methods. In this demonstration, we use the Static method as the proposed algorithm, and also assume that the variant sheep receives only the force of separation \addd{(i.e., $\alpha = (1, 0, 0, 0)$)}.
% \resolved{ハテナを埋めてください．僕がやったら間違うと思うので．}\anna{埋めました}\masaki{thanks}
We show the timeline of the guidance by the FAT method and the proposed method in Figure~\ref{animals}. In Figure~\ref{fig:fat_failure}, we illustrate a typical situation in which the FAT method fails to guide the whole flock because the shepherd keeps trying to guide a variant sheep. On the other hand, as we can see from Figure~\ref{fig:proposed_success}, the proposed method
% \masaki{Static or Dynamic?}\masaki{Staticである}\masaki{反映済み}
enables the shepherd to discriminate normal and variant sheep and to guide the normal sheep successfully in the goal region.

\begin{figure}[tb]
  \centering
  \subfloat[The variant sheep receive one force. ]{\includegraphics[width=.52\linewidth]{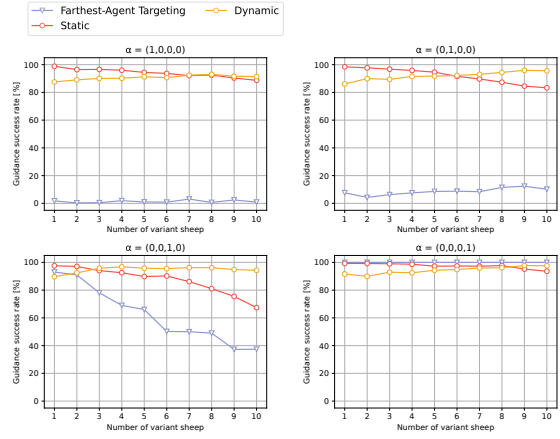}
  \label{fig:virtual_ssn_3}}
  \\
  \subfloat[The variant sheep receive two forces.]{\includegraphics[width=.65\linewidth]{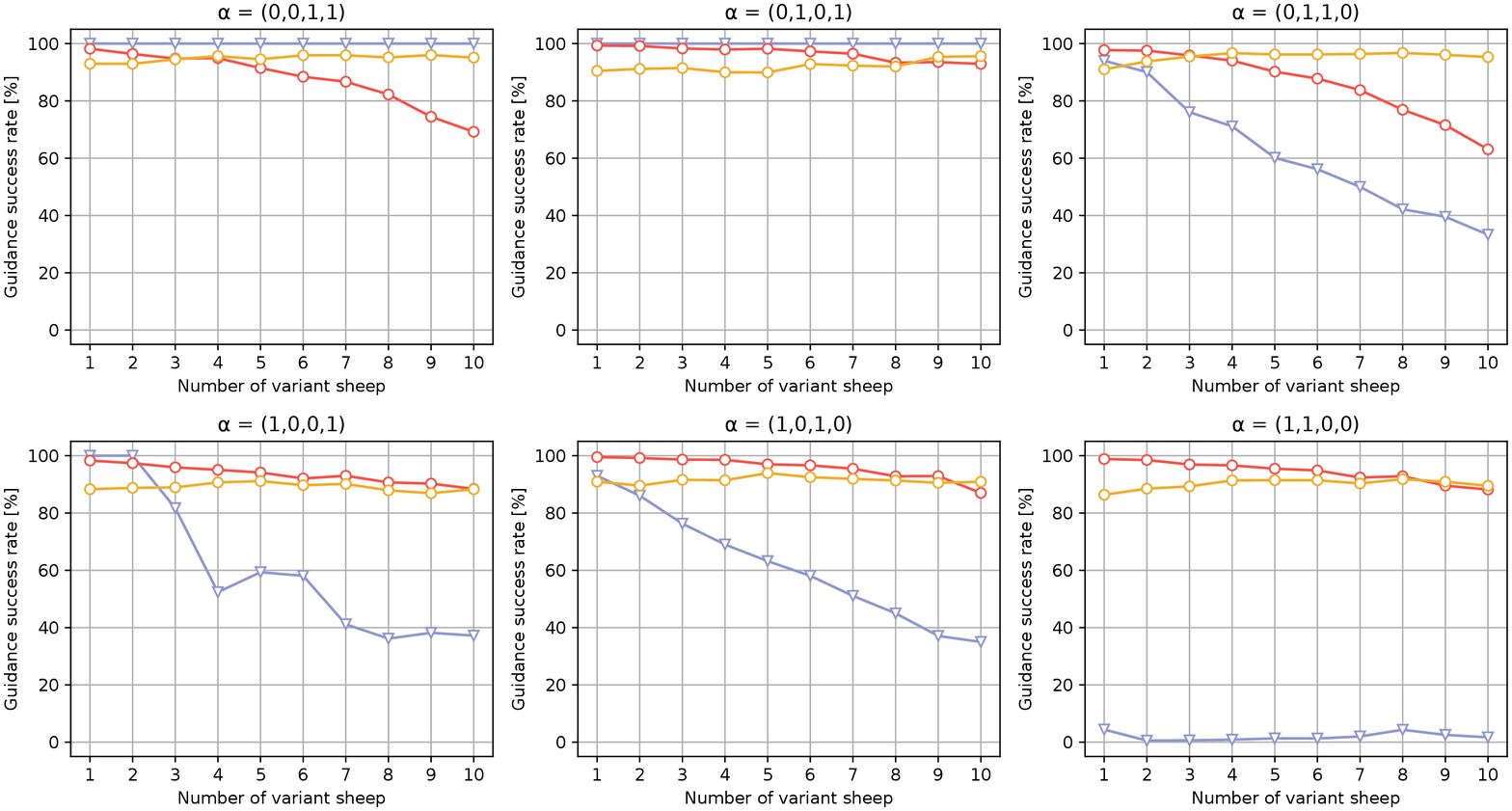}
  \label{fig:virtual_ssn_2}}
  \\
  \subfloat[The variant sheep receive three forces.]{\includegraphics[width=.52\linewidth]{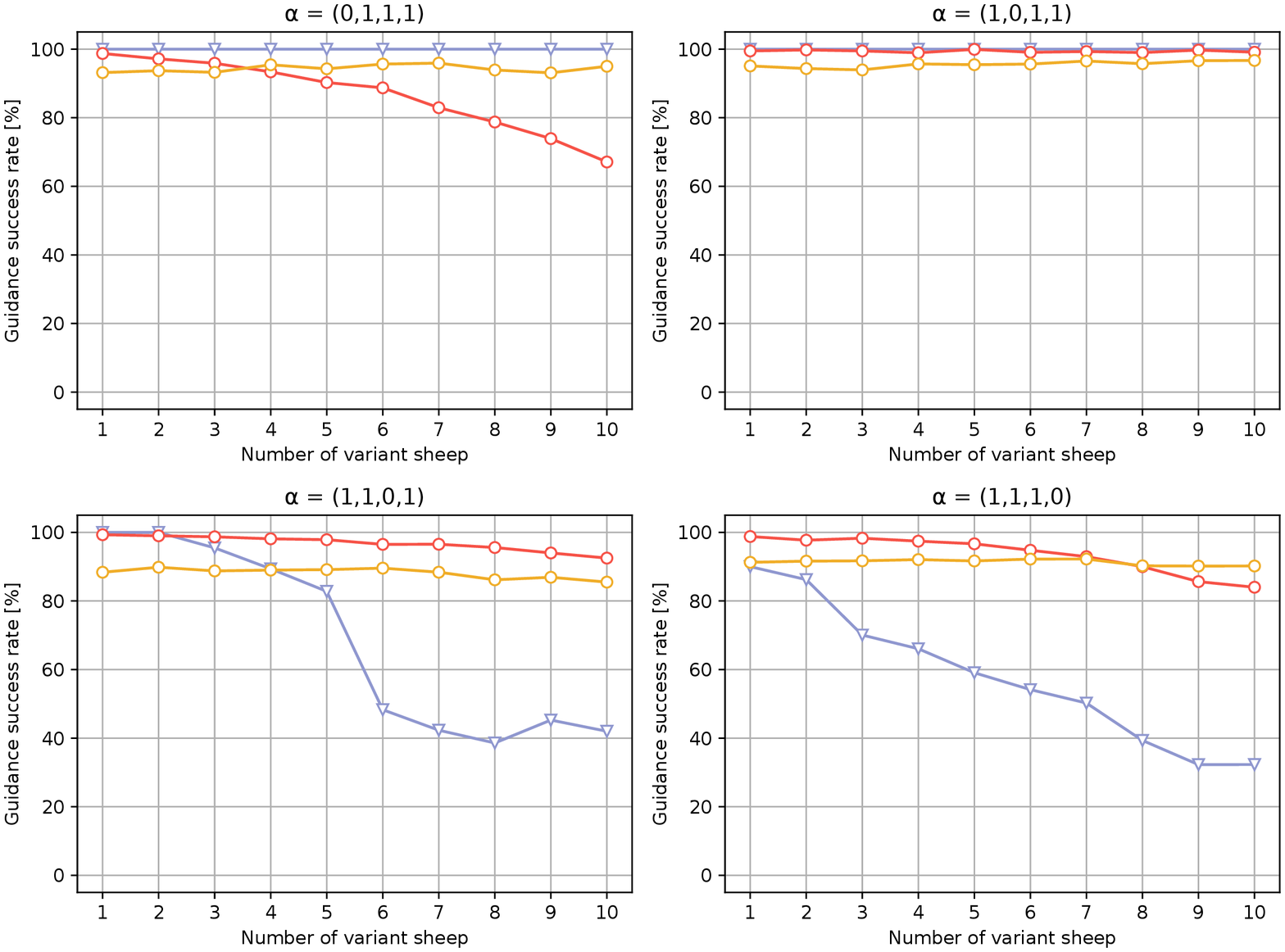}
  \label{fig:virtual_ssn_1}}
%   \caption{従来手法と提案手法による誘導率．横軸：異種の羊の個体数．縦軸：誘導率．\masaki{要英訳}}
  \caption{
%   \masaki{2 columnsになった時に大丈夫か？}\masaki{とりあえずこの図をもって投稿することで合意}
  Guidance success rate by FAT method and two proposed shepherding methods. The blue, red, and orange lines indicate the induction rate by FAT, Static, and Virtual, respectively.
%   \anna{横軸，縦軸の説明していましたが，英訳する段階で，他の論文ではあまり軸の説明がされていないと感じました．ですので，軸の説明から線の色の説明に変えました．でも説明が重複しているような気もしています．確認お願いします．}\masaki{概ねOKです}
  }
  \label{fig:virtual_ssn}
\end{figure}
% \subfloat[異種の羊が○つの力を受ける場合]

\begin{figure}[tb]
  \centering
  \subfloat[The variant sheep receive one force.]{\includegraphics[width=.52\linewidth]{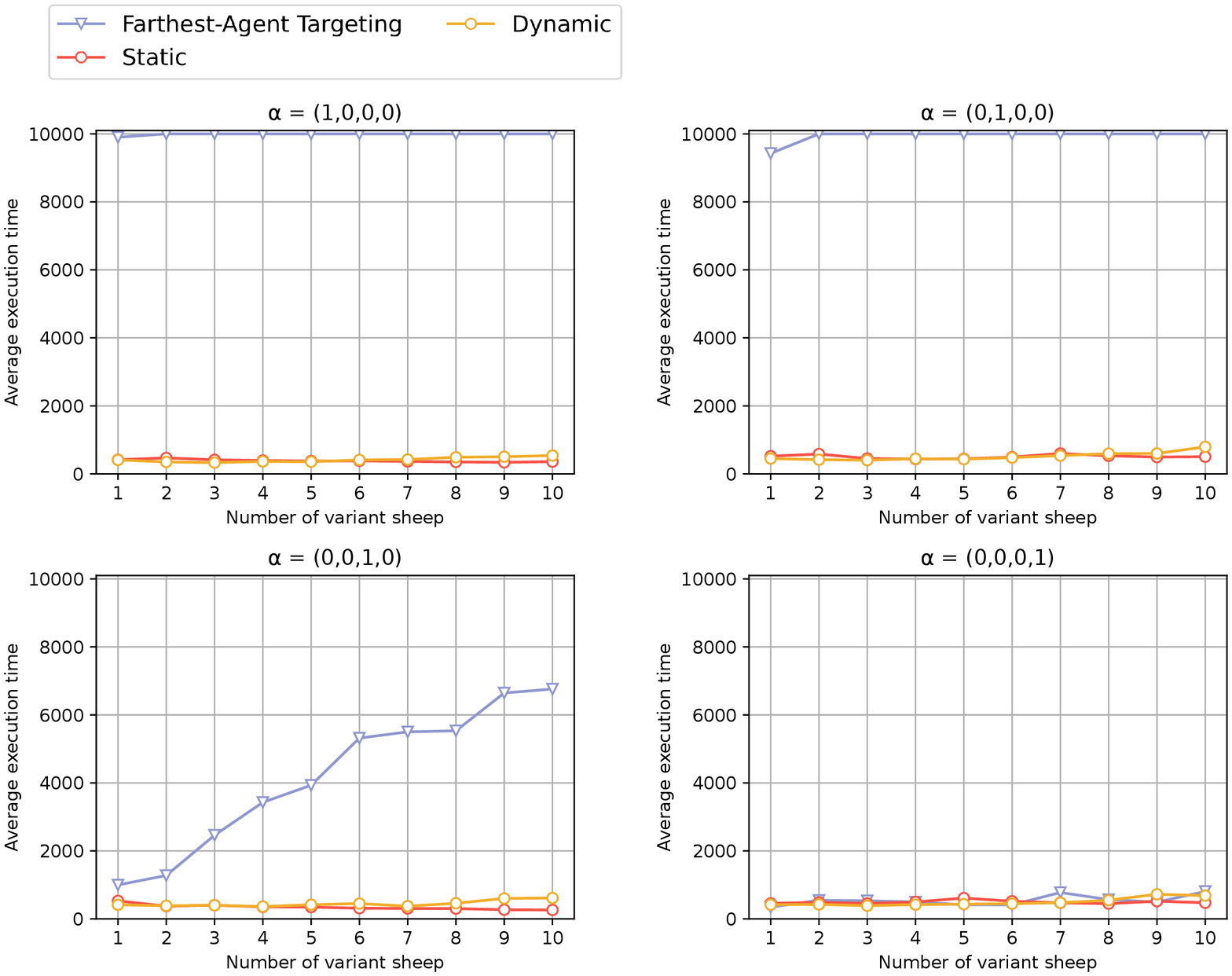}
  \label{fig:virtual_time_3}}
  \\
  \subfloat[The variant sheep receive two forces.]{\includegraphics[width=.65\linewidth]{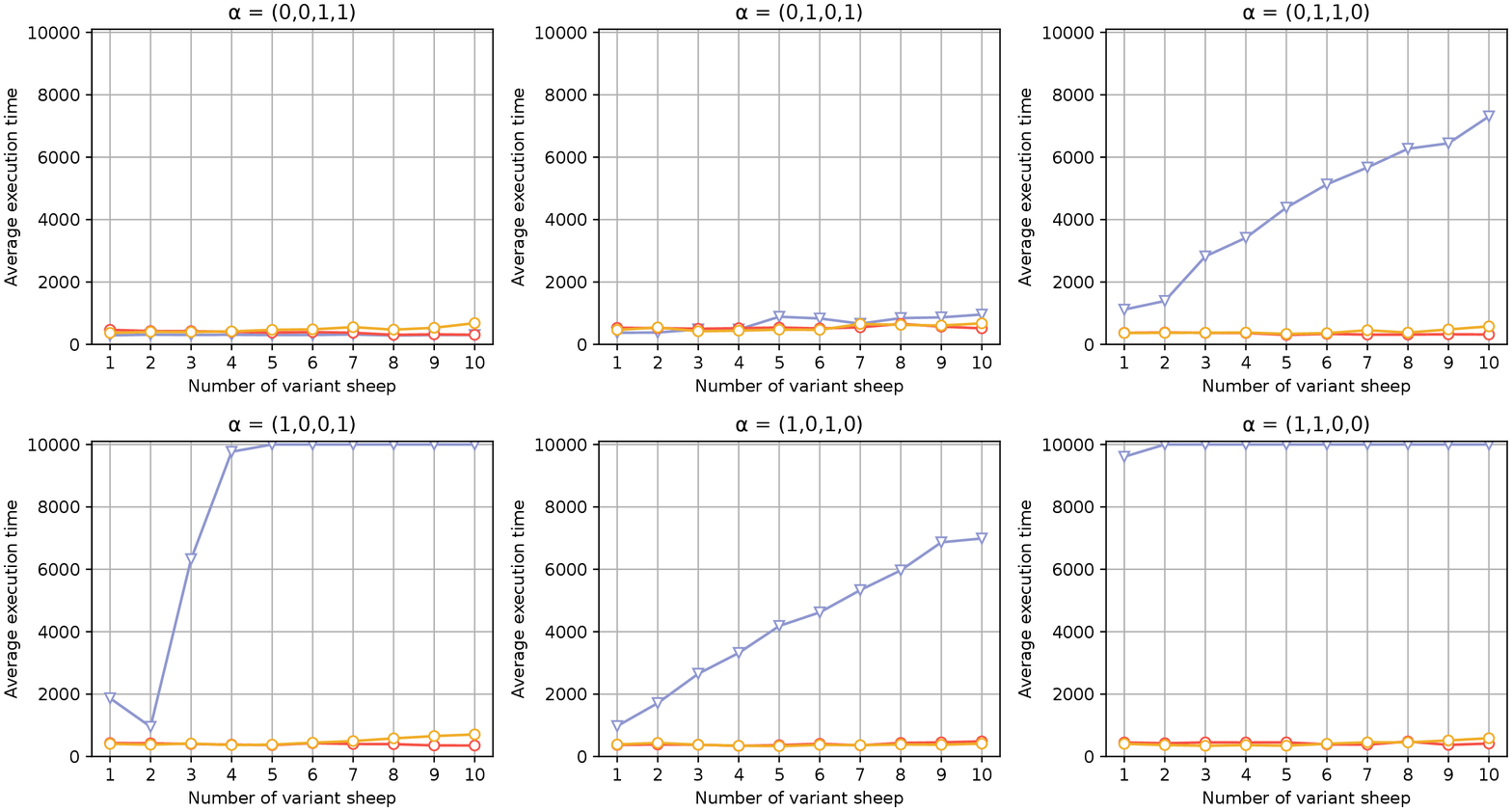}
  \label{fig:virtual_time_2}}
  \\
  \subfloat[The variant sheep receive three forces.]{\includegraphics[width=.52\linewidth]{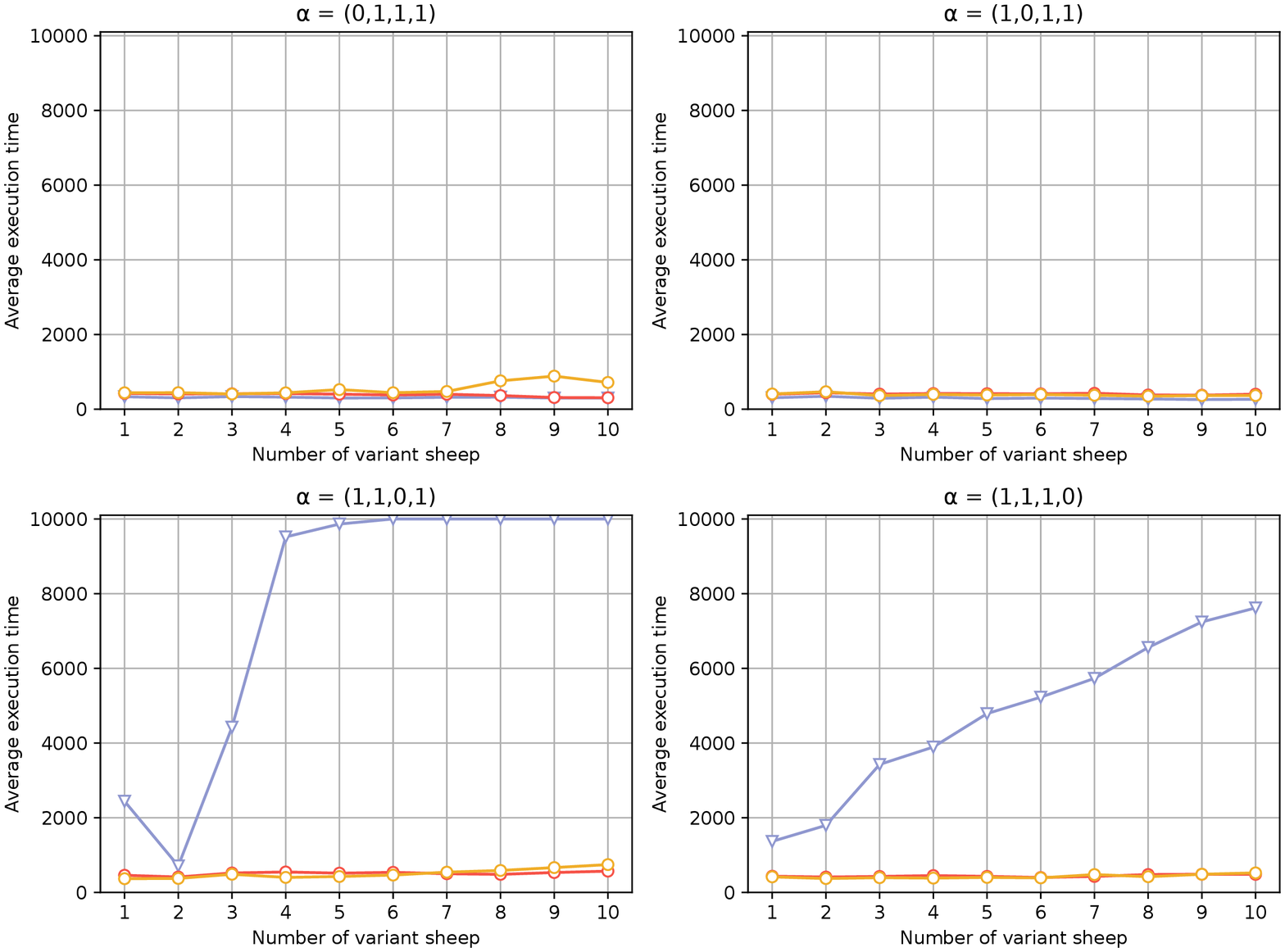}
  \label{fig:virtual_time_1}}
  \caption{Average execution time by FAT method and two proposed shepherding methods. The blue, red, and orange lines indicate the average execution time by FAT, Static, and Virtual, respectively.
%   \masaki{要英訳}\anna{確認お願いします}\masaki{thanks}
}
  \label{fig:time}
\end{figure}

Let us then compare the performances of the proposed methods and the FAT
% \masaki{これをFATと略すかどうかの判断}\masaki{原稿が出来上がって時間があれば検討}\anna{個人的にはFATでいい気がします．論文のルールを知らないので何も言えないですが．}\resolved{意見ありがとう．でしたらFATにしましょう．更新おねがいします}\anna{2回目以降の登場全てを変えました．1章で出てきた後間が空いて3章で再び出てくるのですが，再登場の際一度``Farthest-Agent Targeting''とした方が良いでしょうか？}\masaki{thanks．確かに再登場のときに正式名称出してあげるのも親切だけど，今回はやらなくてもいいと思います．}\anna{わかりました．FATのままにします}
method.
% We define the guidance rate as ...
% \masaki{Guidance rateの定義どこだっけ？}\anna{guidance rate のことでしたら，4.1節最後の段落です}\masaki{thanks}
In Figure~\ref{fig:virtual_ssn}, we show the guidance success rates by the proposed methods (Static and Dynamic) and the FAT method for various values of $M$ (i.e., the number of the variant sheep). The FAT method achieves 100\% guidance success rate for the cases of 1) the variant sheep receives repulsion but not separation ($\alpha = (0, 1, 1, 1),\ (0, 0, 1, 1),\ (0, 1, 0, 1),\ (0, 0, 0, 1)$) and 2) receives all forces but alignment ($\alpha = (1, 0, 1, 1)$). However, for the flock containing variant sheep that receive neither attraction nor repulsion (i.e., when $\alpha = (1, 1, 0, 0),\ (1, 0, 0, 0),\ (0, 1, 0, 0)$), the FAT method frequently fails to guide the flock (guidance success rate $<13\%$). Furthermore, for the case of other types of variant sheep, the FAT method shows the trend in which the guidance success rate decreases with respect to the number of the variant sheep in the flock. On the other hand, we can observe that both of the proposed methods exhibits relatively high performance (guidance success rates $>63\%$) irrespective \dell{for}\adddd{of} the type of variant sheep, confirming their  effectiveness and robustness.
% 二種類の提案手法（Static，Dynamic）およびFarthest-Agent Targeting 法による誘導率をFigure~\ref{fig:virtual_ssn}に示す．これらの図において，横軸は異種の羊の個体数を，縦軸は誘導率を示す．Farthest-Agent Targeting 法では牧羊犬からの斥力を受けかつ分離の力を受けない場合（$\alpha = (0, 1, 1, 1),\ (0, 0, 1, 1),\ (0, 1, 0, 1),\ (0, 0, 0, 1)$）に100\%の誘導率を示す．また，分離，結合，斥力の力を受ける場合（$\alpha = (1, 0, 1, 1)$）にも100\%の誘導率を示す．一方で異種の羊が結合と斥力の力をどちらも受けない（$\alpha = (1, 1, 0, 0),\ (1, 0, 0, 0),\ (0, 1, 0, 0)$）場合には，全ての$M$に対して誘導率が0\%に近い（0.3\% -- 12.4\%）．また，その他の種類の異種の羊の場合には，異種の羊の個体数が増加するにつれて誘導率が減少する傾向が見られる．一方で二種類の提案手法は，ほとんどの場合でFarthest-Agent Targeting 法よりも高い誘導率を示す．提案手法による誘導率は最低でも63.1\%であった．

% \anna{添削\&英訳お願いします．書く事なくて困っています}\masaki{誘導時間が十分に小さい（FATより小さい）ことだけ言っていれば十分かなと思います．英訳について了解です．}
{In order to further investigate the difference in the performances of the proposed and the FAT methods, we compare the average execution time\resolved{pdfで提案してくれたようにfigureの縦軸の変更おねがいします}\anna{しました}\masaki{thanks} (i.e., the average number of steps taken by the algorithms). We show the average execution times of the algorithms in Figure~\ref{fig:time}. We can confirm that the proposed methods finish guidance relatively quickly (average execution time $< 530$). On the other hand, the FAT method requires much longer execution time. This is mainly because the termination criterion of the FAT method is that all the sheep lie in the goal region, which does not often happen when the flock of the sheep contains variant ones.}
% \resolved{とりあえずの英訳です．内容が正しいかどうかの確認おねがいします}\anna{内容確認しました．}\masaki{thanks}
% に示す．これらの図において，横軸は異種の羊の個体数を，縦軸は誘導終了までにかかった時間の平均値を示す．ここで，本報告では，シミュレーションの最大ステップを10000に設定した\masaki{たぶん10000が上限であることを思い出させたいと推測}\masaki{推測通りだった}．

% Farthest-Agent Targeting法では，Figure~\ref{fig:virtual_ssn}において誘導率が低い場合（$\alpha = (0, 1, 1, 1),\ (0, 0, 1, 1),\ (0, 1, 0, 1),\ (0, 0, 0, 1),\ (1, 0, 1, 1)$）に，誘導時間が長くなっていることがわかる．それに対して二つの提案手法（Static，Dynamic）では，全ての場合で誘導時間が600を下回っている（329.89 -- 521.77）．\masaki{Figure~\ref{fig:virtual_ssn}とからめた説明の必要性？}\masaki{からめる必要はない．FATだけ詳しく語るのはヘン．誘導率で勝っているけど時間でめちゃくちゃ負けているわけではない，ということを言えば十分．}

\subsubsection{Comparison of Static and Dynamic methods}\label{subsubsc:pro-dousi}

% 提案する二つの手法の誘導率を比較する．まず，Figure~\ref{fig:virtual_ssn}に示した誘導率を比較する．Static手法では異種の羊の個体
% 数が多くなるほど誘導率が下がる傾向がみられるのに対して，Dynamic手法ではそのような傾向はみられず，異種の種類によっては誘導率が増加する場合もある．この特徴の理由として，固定された閾値を用いるStatic手法では，異種の羊の個体数が大きくなるほど仮想的な羊と通常の羊の軌道が一致しにくくなることで通常の羊を誤って異種と判断する回数が多くなったと考えられる．一方で動的な閾値を用いるDynamic手法では，異種の羊の個体数が少ない場合，仮想的な羊と実際の羊との距離が小さくなるため閾値が小さくなり，通常の羊を異種であると誤判定する回数が多く，誘導率が低かったと考えられる．他方で，異種の羊の個体数が多くなるにつれ閾値が大きくなり，誤判定の回数が少なくなったため，誘導率が上昇したと考えられる．
{In this subsection, we further investigate and discuss the difference in the performance of the two proposed methods (Static and Dynamic). From Figure~\ref{fig:virtual_ssn}, we find that the guidance success rate of the Static method tends to decrease with respect to the number~$M$ of variant sheep. The reason for this characteristic can be attributed to the fact that, the more the variant sheep, the more deviated the trajectories of the virtual sheep to those of the normal sheep. This quantitative change cannot necessarily be appropriately dealt with by the fixed threshold of the Static method. On the other hand, the guidance success rate by the Dynamic method does not exhibit such trend and, furthermore, even increases with respect to $M$ for some types of variant sheep. A possible reason for this phenomenon is that, in the Dynamic method, when there are few variant sheep, its threshold would become relatively small because the difference of the overall dynamics of the flock of virtual sheep and that of actual sheep is small. This would let the threshold of the Dynamic method relatively small, which then can make it difficult for the shepherd to discriminate variant sheep. Similarly, when there are relatively many variant sheep, the Dynamic method would make its threshold high, which then would prevent the shepherd from misjudging a normal sheep as a variant sheep. These two factors can explain the trend in Figure~\ref{fig:virtual_ssn} in which the Dynamic method does not perform as better as the Static method for a small $M$, but can outperform the Static method for larger $M$.}
% \resolved{僕の妄想入っているかもしれませんが，とりあえずこの段落終わらせたので，内容が正しいかどうか確認をお願いします．}\anna{内容確認しました}\masaki{thanks}

\begin{figure}[tb]
  \centering
  \subfloat[$M = 1$]{\includegraphics[width=.48\linewidth]{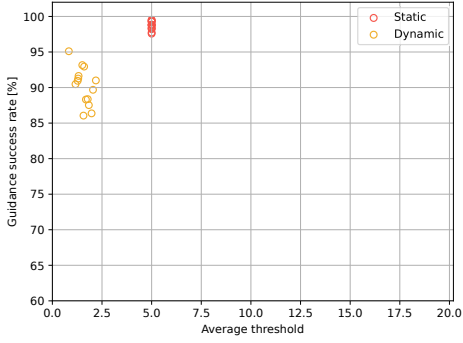}
  \label{fig:shi1}}
  \subfloat[$M = 4$]{\includegraphics[width=.48\linewidth]{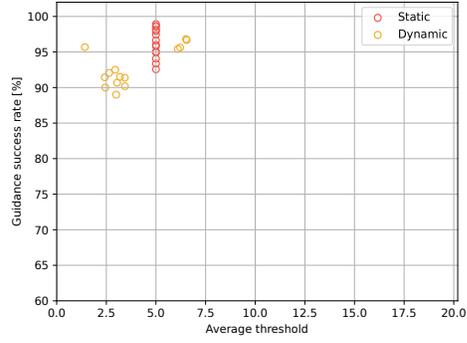}
  \label{fig:shi4}}
  \\
  \subfloat[$M = 7$]{\includegraphics[width=.48\linewidth]{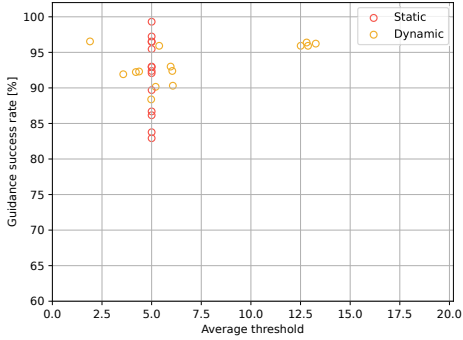}
  \label{fig:shi7}}
  \subfloat[$M = 10$]{\includegraphics[width=.48\linewidth]{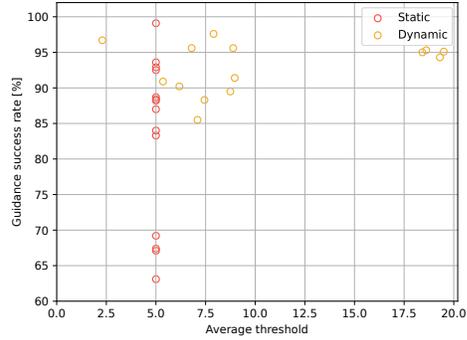}
  \label{fig:shi10}}
%   \caption{閾値の大きさと誘導率の関係．横軸：閾値の大きさ．縦軸：誘導率．\masaki{要英訳}}
  \caption{Relationship between the thresholds and guidance success rate. Horizontal axis: The value of the thresholds. Vertical axis: Guidance success rate.}
  \label{fig:threshold}
\end{figure}

% \anna{ここから書き足した}\masaki{thanks}
% 上述の結果から，誘導率が閾値に依存してどのように変化するか調査した．Figure~\ref{fig:threshold}に閾値の大きさと誘導率の関係を示す．この図において，横軸は閾値の大きさ，縦軸は誘導率を表す．ここで，Static手法における閾値の大きさは5としている．また，Dynamic手法における閾値の大きさは，上述の100回のシミュレーションにおける閾値の大きさの平均値を示す．まずStatic手法では，異種の羊の個体数が増えるにつれて，誘導率が下がりやすい．それに対して，Dynamic手法では誘導率の低下を防いでいる．これは，閾値の大きさを変化させることで，通常の羊を誤って異種と判断する回数がStatic手法よりも少なくなったためであると考えられる．
{In order to further investigate the relationship between the guidance success rate and the thresholds, let us investigate how the guidance success rate depends on the value of the threshold. We show their relationship in Figure~\ref{fig:threshold}.
% In this figure, the horizontal axis represents the threshold value while the vertical axis represents the induction rate.
In the figure, each plot consists of $14\times 2 = 28$ points resulting form all the pairs of the type of the variant sheep and the two threshold method. Therefore, each point represents the average of the threshold and the guidance success rate from the corresponding 100 simulations. The points from the static method lies on the same vertical line because the method uses a fixed threshold~$5$. On the other hand, because the Dynamic method adaptatively changes its threshold in the course of the shepherding guidance, the points do not necessarily lie on the same vertical line.}

\begin{figure}[tb]
\centering
\includegraphics[width=.9\linewidth]{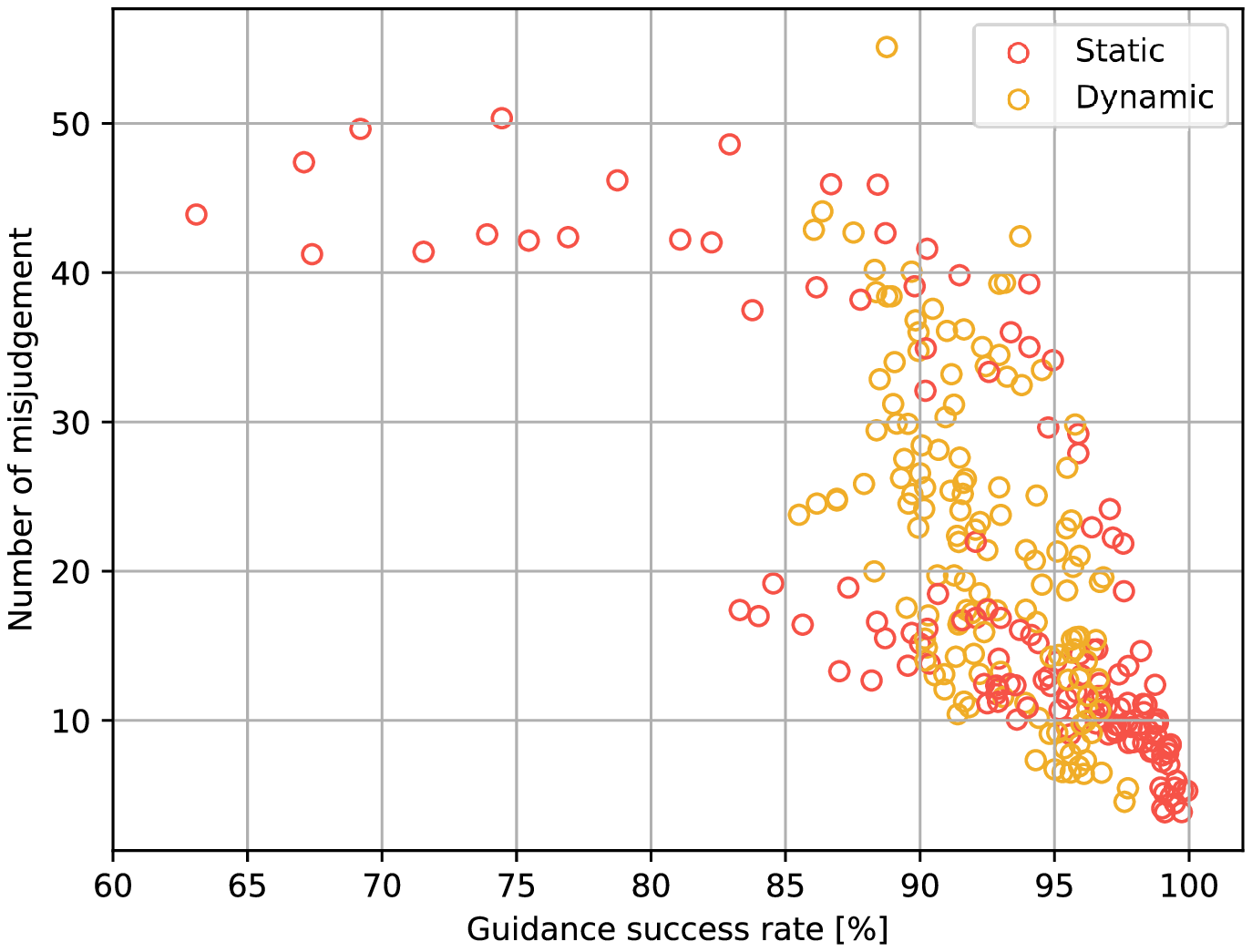}
% \caption{誘導率と誤判定の関係．横軸：誘導率．縦軸：通常の羊を異種と判断した回数．\masaki{要英訳}}
\caption{Relationship between guidance success rate and the number of incorrect discrimination of the normal sheep. Horizontal axis: Guidance success rate. Vertical axis: The number of times a normal sheep is mistakenly judged to be a variant one.}
\label{fig:mis}
\end{figure}

{From Figure~\ref{fig:threshold}, we reconfirm that increasing $M$ can result in degradation of the performance of the Static method. On the other hand, we observe that the average threshold in the Dynamic method tends to increase with $M$, while the guidance success rate is mostly maintained for all the types of the variant sheep. This trend would be because varying the size of the threshold enabled the Dynamic method to reduce the number of times a normal sheep is mistakenly judged to be variant. These observations suggest that preventing a wrong discrimination of a normal sheep would lead to better guidance success rate.}
% \resolved{Figure~\ref{fig:mis}で``guidance success rate''となっています．本文ではおそらくdeeplの都合で``guidance rate''になっています．なので藤岡さん案を尊重して本文で``guidance success rate''としようと思います．いかがでしょうか．}\anna{guidance success rate の方が好きです．こちらでお願いします．現段階のguidance rate は guidance success rate に直しました．}\masaki{了解＆ありがとう}

% このように，異種の羊を正しく検出することが誘導率の向上につながると考えられる．通常の羊を誤って異種と判断した回数と誘導率の関係を調査した．Figure~\ref{fig:mis}に誘導率と誤判定の関係を示す．この図において，横軸は誘導率，縦軸は誤判定の回数を表す．Figure~\ref{fig:mis}より，誘導率と誤判定の回数には負の相関が見られる．よって，上述の仮説の妥当性が示唆される．
% 上記の観察より，異種の羊を正しく検出することが誘導率の向上につながるとの仮説をたてることができる．この仮説を検証するために，通常の羊を誤って異種と判断した回数と誘導率の関係を調べる．我々はFigure~\ref{fig:mis}に誘導率と誤判定回数の関係を示す．この図において，横軸は誘導率，縦軸は誤判定回数を表す．Figure~\ref{fig:mis}より，誘導率と誤判定の回数には負の相関が見られる．よって，上述の仮説の妥当性が示唆される．
{In order to assess the validity of our hypothesis that preventing discrimination of a normal sheep as a variant sheep would lead to increase in the guidance success rate, let us  examine how the guidance success rate depends on the occurrence of the incorrect discrimination. In Figure~\ref{fig:mis}, we show the  relationship between the guidance success rate and the number of incorrect discrimination of the normal sheep. The overall negative correlation from the plot, both in the Static and Dynamic method, confirms the validity of our hypothesis.}
% \resolved{ここも内容の正しさ確認おねがいします．僕はあまり読み返していないので，大きく間違っている可能性があります．}\anna{内容確認しました．}\masaki{thanks}

\section{Conclusion}\label{sc:conclusion}
% \masaki{藤岡さん：内容をもう少し持たせて下さい}\anna{こんな感じでいかがでしょうか．}\anna{本報告では，異種エージェントを含む羊群のshepherding制御のための牧羊犬の機動法則を提案した．牧羊犬は与えられた通常の羊の係数の推定値を基に羊の軌道を予測する．予測軌道から大きく外れた羊を異種の羊と判断し，通常の羊のみをFarthest-Agent Targeting 法により誘導する．本稿では，異種と判断する基準について二つの手法を提案した．シミュレーションの結果，ほとんど全ての場合においてFarthest-Agent Targeting法よりも二つの提案手法の方が高い誘導率を示した．二つの提案手法においては，閾値を動的に変化させるDynamic手法は，異種の羊の個体数が多い場合にも誘導率が低下しなかった．}\resolved{とりあえずで訳してみました．確認おねがいします．}\anna{確認しました．}\masaki{thanks}
{In this paper, we have formulated a shepherding problem for a heterogeneous flock consisting of normal and variant sheep, and then proposed a movement algorithm of the shepherd to solve the formulated shepherding problem. In this algorithm, the shepherd predicts the sheep's trajectories using the given and estimated dynamical model of normal sheep. Then, the shepherd discriminates those sheep deviating from the predicted trajectory as variant. To the agents discriminated to be normal, the shepherd performs navigation control by using the FAT algorithm. We specifically proposed two \dell{algorithms}\adddd{methods} having a different discrimination process; Static, in which the distance threshold for discrimination is constant, and Dynamic, in which the threshold adaptatively changes. Our numerical simulations show that both \dell{algorithms}\adddd{methods} outperform the FAT \dell{algorithm}\adddd{method} in its original form. We also find that the Dynamic \dell{algorithm}\adddd{method} is robust to the change in the number of the variant sheep in the flock of agents. }

% \anna{今後の課題？}\masaki{藤岡さん，日本語で構いませんので今後の課題を述べてください}
% ・異種の種類を増やす
% ・牧羊犬が分かる情報かえる　通常羊の係数推定値　誤差の範囲，どれがかわりそう
% ・ほかの手法
% \anna{今後の課題として，同時に存在する異種の種類を複数とした場合の誘導率の調査が挙げられる．また，本稿ではFarthest-Agent Targeting法を基に手法を提案したが，他の従来法において，提案した異種判別のアルゴリズムを適用した場合の誘導率を調査したい．さらに，牧羊犬の持つ情報を追加（例えば，係数の推定値の誤差の範囲や，欠如しやすい力の種類）した場合における誘導アルゴリズムの設計を今後の課題とする．}\resolved{正しさの確認お願いします．書いてくれた三個目は分量の関係で除外しました．}\anna{確認しました．}\masaki{thanks}
{There are several interesting directions of future research. One is a further comprehensive investigation of the performance of the proposed method. For example, in this paper, we have focused on the situation in which the uniformity among the variant sheep is guaranteed. Investigating how the heterogeneity within the variant sheep affects the performance of the proposed methods is necessary to further establish their effectiveness. Another direction is to validate the effectiveness of the proposed discrimination \dell{algorithm}\adddd{method} when used in the shepherding \dell{algorithms}\adddd{methods} other than the FAT \dell{algorithms}\adddd{method}.}

% \masaki{Referencesは最後の最後にbblファイルの内容をコピペした上で調整しましょう．}
% \bibliographystyle{tfnlm}
% \bibliography{references,library}

\end{document}